\documentclass[aps,prd,preprint,preprintnumbers,amsmath,amssymb,groupedaddress,nofootinbib]{revtex4}

\usepackage[]{graphicx}
\usepackage{amscd}
\usepackage[all,cmtip]{xy}
\usepackage{mathrsfs}
\usepackage[margin=10pt,font=small,labelfont=bf]{caption}
\usepackage{simplewick}
\usepackage{changepage}
\usepackage{bm}

\newcommand{\bea}{\begin{eqnarray}}
\newcommand{\eea}{\end{eqnarray}}
\newcommand{\beq}{\begin{equation}}
\newcommand{\eeq}{\end{equation}}

\newcommand{\tl}{\tilde}

\newcommand{\De}{\Delta}
\newcommand{\la}{\lambda}
\newcommand{\de}{\delta}

\def\beq{\begin{equation}}
\def\eeq{\end{equation}}
\def\<{\langle}
\def\>{\rangle}

\def\cO {{\cal O}}

\def\thex{\theta_{\rm ext}}
\def\bthex{\bar{\theta}_{\rm ext}}
\def\cG {{\cal G}}

\def\cN {{\cal N}}

\def\cS {{\cal S}}

\def\cX {{\cal X}}
\def\cY {{\cal Y}}
\def\cW {{\cal W}}
\def\cS {{\cal S}}
\def\cT {{\cal T}}

\newcommand{\mC}{\mathbb{C}}

\begin{document}

\preprint{MI-TH-1607}

\title{The Most General $4\mathcal{D}$ $\mathcal{N}=1$ Superconformal Blocks for Scalar Operators}

\author{ \vspace{8mm}Zhijin Li}
\email{lizhijin@physics.tamu.edu}
\author{Ning Su}
\email{suning1985@gmail.com}
\affiliation{{\small \vspace{4mm} George P. and Cynthia W. Mitchell Institute for
Fundamental Physics and Astronomy,
Texas A\&M University, College Station, TX 77843, USA} \vspace{16mm}}

\begin{abstract}
\vspace{3mm}
  We compute the most general superconformal blocks for scalar operators in $4\mathcal{D}$ $\mathcal{N}=1$ superconformal field theories. Specifically we employ the supershadow formalism to study the four-point correlator $\<\Phi_1 \Phi_2 \Phi_3 \Phi_4\>$, in which the four scalars $\Phi_i$ have arbitrary scaling dimensions and R-charges with the only constraint from R-symmetry invariance of the four-point function. The exchanged operators can have arbitrary R-charges. Our results extend previous studies on $4\mathcal{D}$ $\mathcal{N}=1$ superconmformal blocks to the most general case, which are the essential ingredient for superconformal bootstrap, especially for bootstrapping mixed correlators of scalars with independent scaling dimensions and R-charges.
\end{abstract}

\maketitle

\newpage

\tableofcontents

\newpage

\section{Introduction}
The conformal bootstrap program, which was initially proposed for two dimensional conformal field theories (CFTs )\cite{Ferrara:1973yt, Polyakov:1974gs, Mack:1975jr} has been found to be remarkably powerful to study CFTs in higher dimensional spacetime  \cite{Rattazzi:2008pe}. The crossing symmetry and unitarity condition  can provide strong constraints on the operator scaling dimensions, coefficients in operator product expansion (OPE) and the central charges \cite{Rychkov:2009ij,Poland:2010wg,Poland:2011ey,ElShowk:2012ht,Kos:2013tga,Beem:2013qxa,Beem:2014zpa,El-Showk:2014dwa,
Kos:2014bka,Chester:2014fya,Chester:2014gqa,Bae:2014hia,Bobev:2015jxa,Kos:2015mba,Chester:2015qca,Beem:2015aoa,Iliesiu:2015qra,
Lemos:2015awa,Lin:2015wcg,Chester:2015lej,Li:2015itl}. The most striking results are obtained in \cite{Kos:2014bka,Kos:2015mba}, in which the classical $3D$ Ising model and $O(N)$ vector model are studied through bootstrapping the mixed correlators. It is shown that by imposing certain reasonable assumptions on the spectrum, the CFT data  can be isolated in small islands. These results are expected to be generalized to the supersymmetric theories, in which supersymmetry provides strong constraints on the quantum dynamics and leads to abundant conformal theories. The supersymmetric conformal bootstrap is especially important for $4\mathcal{D}$ theories since most of the known $4\mathcal{D}$ CFTs are of supersymmetric conformal field theories (SCFTs).

The critical ingredient utilized in conformal bootstrap is the convexity of conformal blocks \cite{Rattazzi:2008pe}. The four-point functions can be decomposed into conformal partial waves which describe the exchange of primary operators together with their descendants. As for the SCFTs, it can be shown from superconfomral algebra that a superconformal primary multiplet can be decomposed into (finite) many conformal primary multiplets, consequently the superconformal block is the summation of several conformal blocks with coefficients restricted by supersymmetry. Previous results on $4\mathcal{D}$ supercomformal blocks have been presented in \cite{Poland:2010wg,Dolan:2001tt,Dolan:2004iy,Dolan:2004mu,Fortin:2011nq,Berkooz:2014yda} based on the superconformal Casimir approach. These studies are mainly focused on the four-point functions of chiral-antichiral fields or conserved currents, which are protected by short-conditions or symmetries. For the four-point functions of more general fields, the traditional superconformal Casimir approach becomes less helpful due to the complex superconformal invariants appearing in the superconformal blocks.
Recently a new covariant approach based on the supershadow formalism has been proposed in \cite{Fitzpatrick:2014oza} and applied in \cite{Khandker:2014mpa} for $\mathcal{N}=1$ superconformal blocks corresponding to exchange of operators neutral under the $U(1)_R$ symmetry.

The new covariant approach generalizes the embedding and shadow formalisms proposed for CFTs to treat with supersymmetric theories. The embedding formalism
\cite{Dirac:1936fq,Mack:1969rr,Ferrara:1973eg,Weinberg:2010fx,Costa:2011mg,Costa:2011dw,SimmonsDuffin:2012uy}
realizes conformal transformations linearly and provides a convenient way to construct conformally covariant correlation functions.
Specifically, the conformal covariance of correlation function is mapped into Lorentz covariance of the correlation function in embedding space.
Recently the embedding formalism has been widely used to study the conformal blocks of spinor or tensor operators \cite{Elkhidir:2014woa, Echeverri:2015rwa, Iliesiu:2015qra,  Rejon-Barrera:2015bpa,Iliesiu:2015akf, Li:2015itl, Echeverri:2016dun}.
The $SU(2,2|\mathcal{N})$ superconformal symmetry transformations can be linearly realized in the supersymmetric generalization--superembedding space \cite{Goldberger:2011yp, Goldberger:2012xb, Siegel:2012di, Maio:2012tx, Kuzenko:2012tb}.
The shadow formalism was first proposed in \cite{Ferrara:1972xe,Ferrara:1972ay,Ferrara:1972uq} and recently applied in computing conformal blocks \cite{SimmonsDuffin:2012uy}. Using the shadow operators one can construct projectors of the four-point function which decomposes the four-point function into
conformal blocks represented by the exchanged primary operator, actually it provides an analytical method to compute the conformal blocks, and similarly, its supersymmetric generalization  gives a systematic method to study the $\mathcal{N}=1$ superconformal blocks.

In this work we will apply the supershadow formalism to study the most general $\mathcal{N}=1$ four-point functions of scalars $\<\Phi_1 \Phi_2 \Phi_3 \Phi_4\>$,
where the scalars $\Phi_i$ have independent scaling dimensions and R-charges. The only constraint is from vanishing net R-charges of four scalars so that the $U(1)_R$ symmetry is preserved. Through partial wave decomposition the four-point function gives rise to the most general superconformal blocks, which provide crucial ingredients for $\mathcal{N}=1$ superconformal bootstrap. Our results are especially important for bootstrapping mixed correlators of scalars with
arbitrary scaling dimensions and R-charges, which are beyond previous results on $\mathcal{N}=1$ superconformal blocks. A rather interesting problem
is to bootstrap the mixed correlators between chiral and real scalars which appear in the minimal $4\mathcal{D}$ $\mathcal{N}=1$ SCFT \cite{Poland:2011ey, Poland:2015mta, Li:toa}.

This paper is organized as follows. In section 2 we briefly review the superembedding space, supershadow formalism and their roles in computing $\cN=1$ superconformal partial waves.
In section 3 we study the most general three-point correlators consisting of two scalars and a spin-$\ell$ operator with arbitrary scaling dimensions and $U(1)$ R-charges. In section 4 we compute the superconformal partial waves, which are the supershadow projection of the four-point function and obtained from products of two three-point functions.
The major difficulty comes from different superconformal weights of scalars, which break the symmetry under  coordinate interchange $1\leftrightarrow3$, $2\leftrightarrow4$. Without such symmetry it gets more tricky to evaluate the superconformal integragtions.
We present the final results on superconformal blocks in section 5, and compare our general superconformal blocks with known examples as a non-trivial consistent check. Conclusions are made in section 6.
We will follow the conventions used in \cite{Fitzpatrick:2014oza,Khandker:2014mpa} throughout this paper.

\section{Brief Review of Superembedding Space and Supershadow Formalism}
In this part we briefly review the superembedding space and supershadow formalism, especially for the techniques needed in our computation. More details on these topics are presented in  \cite{Goldberger:2011yp, Goldberger:2012xb, SimmonsDuffin:2012uy, Fitzpatrick:2014oza, Khandker:2014mpa}.
\subsection{Superembedding Space}
There are two equivalent ways to construct superspace in which the $4\mathcal{D}$ $\mathcal{N}=1$ superconformal group $SU(2,2|1)$ acts linearly. A natural choice is to construct (anti-) fundamental representation of $SU(2,2|1)$, the (dual) supertwistor $\cY_A\in \mC^{4|1}$ ($\bar\cY^A$):
\beq
\cY_A =
\begin{pmatrix}
Y_\alpha\\
Y^{\dot\alpha}\\
Y_5
\end{pmatrix}, ~~~~~~~
\bar \cY^A =
\begin{pmatrix}
\bar Y^\alpha &
\bar Y_{\dot\alpha} &
\bar Y^5
\end{pmatrix},
\eeq
where $Y_\alpha$ and $Y^{\dot\alpha}$ are bosonic complex components while $Y_5$ is fermionic.
Representation for extended supersymmetry $\mathcal{N}>1$ can be realized with more fermionic components in the supertwistors.

The well-known $4\mathcal{D}$ $\mathcal{N}=1$ chiral superspace $(x_+^{\dot\alpha \alpha},\theta_i^\alpha)$ can be reproduced from a pair of supertwistors $\cY^m_i$, $m=1,2$, with following constraints
\beq
\bar \cY^{n A}\cY_A^m = 0,~~~~ m, n = 1,2. \label{sconst}
\eeq
Here one needs to fix the $\rm{GL}(2,\mC)$ gauge redundancy  arising from the rotation of the two supertwistors, and similarly for the dual supertwistors.
Taking the gauge named ``Poincar\'e section", the supertwistor and its dual are simplified into
\beq
\cY_A^m = \begin{pmatrix}
\de_\alpha{}^m\\
ix_+^{\dot\alpha m}\\
2\theta^m
\end{pmatrix},
~~~~~~~
\bar \cY^{n A} = \begin{pmatrix}
-ix_-^{n \alpha} & \de^{n}_{\dot \alpha} & 2\bar\theta^{n}
\end{pmatrix}.
\eeq
In the ``Poincar\'e section" the constraints (\ref{sconst}) turn into $x_+-x_--4i\bar\theta\theta=0$ and can be solved by the
chiral-antichiral coordinates of $4\mathcal{D}$ $\mathcal{N}=1$ superspace.


The superembedding space provides another way to realize superconformal transformations linearly. Its coordinates are bi-supertwistors ($\cX,\bar\cX$)
\beq
\cX_{AB} \equiv \cY^m_A\cY^n_B\epsilon_{mn},\qquad \bar \cX^{AB}\equiv \bar \cY^{i A}\bar \cY^{j B}\epsilon_{ij},
\eeq
By construction, the bi-supertwistors are
invariant under $\rm SL(2,\mC)\times\rm SL(2,\mC)$
and significantly reduce the gauge redundencies of supertwistors, besides, they satisfy the ``null" conditions
\beq
\bar\cX^{AB}\cX_{BC}=0. \label{null}
\eeq
Superconformal invariants are obtained from superstraces of successive products of $\cX$'s and $\bar\cX$'s.
For example, the two-point invariant $\<\bar2 1\>\equiv{\rm Tr}(\bar\cX_2\cX_1)$ \footnote{Here and after the indices $(j,\bar k,\cdots)$ denote the superembedding variables $(\cX_j,\bar\cX_k,\cdots)$.} is
\beq
\<\bar2 1\> \equiv \bar\cX_2^{AB}\cX_{1BA}=-2(x_{2-}-x_{1+}+2i\theta_1\sigma\bar\theta_2)^2,
\eeq
where the last step is evaluated in the Poincar\'e section and it is
easy to show that
\beq
\<\bar2 1\>^\dagger=\<\bar1 2\>.
\eeq

The $\mathcal{N}=1$ superconformal multiplets can be directly lifted to superembedding space.
There are four parameters to characterize a $4\mathcal{D}$ $\mathcal{N}=1$ superconformal primary superfield $\cO$:
the $\rm SL(2,\mC)$ Lorentz quantum numbers $(\frac{\ell}{2},\frac{\bar \ell}{2})$, the scaling dimension $\Delta$ and
$U(1)_R$ charge $R_\cO$. For SCFTs, usually it is more convenient to use superconformal weights $q,\bar q$
\begin{equation}
q \equiv \frac{1}{2} \left(\Delta + \frac{3}{2}R_\cO\right), ~~~~~~ \bar{q} \equiv \frac{1}{2} \left(\Delta - \frac{3}{2}R_\cO\right)
\end{equation}
rather than the scaling dimension $\Delta$.
Given a superfield $\phi_{\alpha_1\cdots\alpha_\ell}^{\dot{\beta}_1\cdots\dot{\beta}_{\bar{\ell}}}: (\frac{\ell}{2},\frac{\bar \ell}{2}, q,\bar q)$, its map in superembedding space is a multi-twistor
$\Phi_{B_1\cdots B_{\bar{\ell}}}^{\phantom{A_{\ell}\cdots A_{1}}A_{1}\cdots A_{\ell}}(\cX,\bar \cX)$ with homogeneity
\beq
\Phi(\la \cX,\bar \la \cX) = \la^{-q-\frac{\ell}{2}}\bar\la^{-\bar q-\frac{\bar \ell}{2}}\Phi(\cX,\bar \cX).
\eeq
The twistor indices make the computations cumbersome, especially for operators with large spin $\ell$. Such difficulty is overcome  
in \cite{Costa:2011mg} based on 
an index-free notation for non-supersymmetric CFTs. The index-free notation is further generalized for $\cN=1$ $4D$ SCFTs in \cite{Fitzpatrick:2014oza}.
The authors introduced pairs of null auxiliary twistors $\cS_A,\bar \cS^A: ~\bar \cS^A\cS_A=0$, which are used to contract with twistor indices of lifted fields
\beq
\Phi(\cX,\bar{\cX},\cS,\bar{\cS})\equiv\bar{\cS}^{B_{\bar{\ell}}}\cdots\bar{\cS}^{B_{1}}\Phi_{B_{1}\cdots B_{\bar{\ell}}}^{\phantom{A_{n}\cdots A_{1}}A_{1}\cdots A_{\ell}}\cS_{A_{\ell}}\cdots \cS_{A_{1}}.
\eeq
As construction, $\Phi(\cX,\bar{\cX},\cS,\bar{\cS})$ is a polynomial of $\cS_A,\bar \cS^A$ while with no tensor index,
and conversely, one can reproduce the initial superfield from the index-free superembedding fields $\Phi(\cX,\bar{\cX},\cS,\bar{\cS})$ through
\beq
\phi_{\alpha_1\cdots\alpha_\ell}^{\dot{\beta}_1\cdots\dot{\beta}_{\bar{\ell}}} = \left.\frac{1}{\ell!}\frac{1}{\bar{\ell}!} \left(\bar{\cX}\overrightarrow{\partial_{\bar{\cS}}}\right)^{\dot{\beta}_1}\cdots\left(\bar{\cX}\overrightarrow{\partial_{\bar{\cS}}}\right)^{\dot{\beta}_{\bar{\ell}}} \Phi(\cX,\bar{\cX},\cS,\bar{\cS}) \left(\overleftarrow{\partial_\cS} \cX\right)_{\alpha_1} \cdots \left(\overleftarrow{\partial_\cS} \cX\right)_{\alpha_\ell} \right|_\textrm{Poincar\'e} . \label{map}
\eeq
To fix gauge redundancies in the lifted fields the auxiliary fields are set to be transverse $\bar{\cX}\cS=0,~\bar{\cS}\cX=0$.  

Strings with auxiliary fields, like $\bar \cS_i j\bar k l\cdots\bar m \cS_n$ are superconformal invariant so provide a new type of  superconformal invariants besides the supertraces of superembedding coordinates.  Correlation functions are built from
the two kinds of superconformal invariants. In particular, the two-point function can be completely determined by imposing homogeneity conditions.

It gets more tricky in evaluating three-point functions $\<\Phi_1(1,\bar 1)\Phi_2(2,\bar 2)\Phi_3(3,\bar 3)\>$. For nonsupersymmetric CFTs, conformal symmetry and homogeneities of lifted fields are sufficient to fix three-point functions up to a constant.
While for SCFTs, the degree of freedoms of superembedding coordinates are notably enlarged by fermionic components, and
it is possible to construct superconformal invariant cross ratio even for three-point correlator, in contrast in CFTs it is impossible to construct conformal
invariant cross ratio with fields less than $4$. 
The invariant cross ratio is built from supertraces \cite{Park:1997bq,Park:1999pd,Goldberger:2012xb}
\beq
u=\frac{\langle 1\bar{2}\rangle \langle 2\bar{3}\rangle \langle 3\bar{1}\rangle}{\langle 2\bar{1}\rangle \langle 3\bar{2}\rangle \langle 1\bar{3}\rangle},
\eeq
which has no contribution on the homogeneity. In consequence, the three-point function can be arbitrary function of the cross ratio $u$.
 Denoting
\beq
z=\frac{1-u}{1+u},
\eeq
one can show that  $z$ is proportional to the fermionic components $\theta_i,~ \bar\theta_i$ and satisfies
\beq
z^{3}=0,~~~~~z|_{1\leftrightarrow2}=z^\dagger=-z.
\eeq
Therefore the most general function of $z$ appearing in the three-point function is up to the second order, besides, considering its symmetry property under permutation $1\leftrightarrow2$, there are four free parameters in the general three-point functions \cite{Khandker:2014mpa}. Additional restrictions, like chirality can provide strong constraints on the parameters and simplify the three-point functions drastically. More details on the three-point correlators of general scalars will be studied  in Section 3.

\subsection{Supershadow Formalism}
The supershadow approach is based on the observation that two operators $\cO:(\frac{\ell}{2},\frac{\bar \ell}{2}, q,\bar q)$ and $\tl\cO:(\frac{\bar \ell}{2},\frac{ \ell}{2}, 1-q,1-\bar q)$ share the same superconformal Casimir so have non-vanishing two-point function.  Then the operator $\tl\cO$, which is referred to shadow operator of $\cO$, can be used to project the correlation functions onto irreducible representation of $\cO$, i.e., the superconformal partial wave corresponding to exchange primary field $\cO$ and its descendants.

The shadow operator $\tl\cO$ can be constructed from $\cO$ through
\beq
\tl{\mathcal{O}}(1,\bar{1},\cS,\bar{\cS})\equiv\int D[2,\bar{2}]\frac{{\mathcal{O}^\dag}(2,\bar2,2\bar{\cS},\bar{2}\cS)}{\langle 1\bar{2}\rangle ^{1-q+\frac{\ell}{2}}\langle \bar{1}2\rangle ^{1-\bar{q}+\frac{\bar{\ell}}{2}}},
\label{shadow}
\eeq
where $D[2,\bar{2}]$ gives the superconformal measure. One can show that the operator obtained from (\ref{shadow}) has the expected quantum numbers of shadow operator $\tl\cO$. Then it is straightforward to write down the projector
\begin{equation}
\left|\mathcal{O}\right|=  \frac{1}{\ell!^{2}\bar{\ell}!^{2}}\int_M D[1,\bar{1}]\mathcal{O}(1,\bar{1},\cS,\bar{\cS})\rangle \left(\overleftarrow{\partial_{\cS}}1\overrightarrow{\partial_{\cT}}\right)^{\ell}\left(\overleftarrow{\partial_{\bar{\cS}}}\bar{1}\overrightarrow{\partial_{\bar{\cT}}}\right)^{\bar{\ell}}\langle \tl{\mathcal{O}}(1,\bar{1},\cT,\bar{\cT}) \hspace{2mm},
\label{Projector}
\end{equation}
in which the denotation $M$ indicates ``monodromy projection" \cite{SimmonsDuffin:2012uy}. By inserting the projector $\left|\mathcal{O}\right|$ into the four-point function $\<\Phi_1\Phi_2\Phi_3\Phi_4\>$ one can get the superconformal partial wave $\cW_\cO$
\begin{equation}
\mathcal{W_{O}}\propto  \< \Phi_1\Phi_2 \left|\mathcal{O}\right| \Phi_3\Phi_4 \>,
\end{equation}
which corresponds to exchange $\cO$ and its descendants. Here the supershadow projector reduces the four-point function into a product of two three-point functions $\<\Phi_1\Phi_2\cO\>$ and $\<\tl\cO \Phi_3\Phi_4\>$, which as discussed before, can be easily obtained from superembedding formalism.

The remaining problem is to evaluate the integration in superembedding space. Normally the integrations involve in both bosonic and fermionic components and are rather complex, while for the scalar four-point functions, where the external fermionic components of
$\Phi_i$ are vanished $\theta_i\equiv \thex=0$, it was proved in \cite{Fitzpatrick:2014oza} that the integrations can be simplified into non-supersymmetric cases
\beq
\int D[\cY,\bar{\cY}]g(\cX,\bar\cX)|_{\thex=\bthex=0}=\int D^4X\partial^{2}_{\bar X} g(X,\bar X)|_{\bar X=X}, \label{Int1}
\eeq
where the embedding coordinates $X$'s are the bosonic part of superembedding coordinates $\cX$'s. Right hand side integration in embedding space has been comprehensively studied in \cite{SimmonsDuffin:2012uy}.

Combining all these materials together one can study the $\mathcal{N}=1$ superconformal blocks analytically and the results can be expressed in a compact form. Superconformal partial wave $\cW_\cO$ for real ($U(1)_R$ neutral) $\cO$ has been studied in \cite{Khandker:2014mpa}. In the following part we will apply this method to solve the most general superconformal partial waves.

\section{General Three-Point Functions}
In this section we analyze the most general three-point function $\<\Phi_1(1,\bar 1)\Phi_2(2,\bar 2)\cO(0,\bar 0)\>$. The scalars $\Phi_1,~\Phi_2$
have independent superconformal weights $(q_1,\bar q_1)$ and $(q_2,\bar q_2)$, respectively. The exchanged superprimary operator $\cO$ has quantum numbers
$(\frac{\ell}{2},\frac{\ell}{2}, \De, R_{\cO})$, where its $U(1)_R$ charge is $R_{\cO}= \frac{2}{3}R\equiv\frac{2}{3}(\bar q_1+\bar q_2-q_1-q_2)$.
From superembedding coordinates we can construct superconformal invariants $\<i\bar j\>$ with $i,j\in{0,1,2}$, two elementary tensor structures
\beq
S\equiv\frac{\bar{\cS}1\bar{2}\cS}{\langle 1\bar{2}\rangle },~~~~~~ S|_{1\leftrightarrow2}=S^\dagger\equiv\frac{\bar{\cS}2\bar{1}\cS}{\langle 2\bar{1}\rangle},
\eeq
and also the invariant cross ratio $z$. For superprimary operators $\cO$ with spin-$\ell$, it is useful to construct following ``eigen" tensor structures with parity $\pm(-1)^\ell$ under coordinate interchange $1\leftrightarrow2$:
\bea
S_{-}^{\ell} & = & \frac{1}{2}\left(S^{\ell}+(-1)^{\ell}(1\leftrightarrow2)\right),\nonumber \\
S_{+} S_{-}^{\ell-1} & = & \frac{1}{2\ell}\left(S^{\ell}-(-1)^{\ell}(1\leftrightarrow2)\right).
\eea
All the spin-$\ell$ tensor structures $S_{+}^{m} S_{-}^{\ell-m}$ with $m\geqslant2$ vanish due to the null condition of $S_+$.

The most general three-point function is constructed in terms of supertraces, invariant cross ratio and tensor structures as follows:
\bea
\langle \Phi_{1}(1,\bar{1})\Phi_{2}(2,\bar{2})\cO(0,\bar{0},\cS,\bar{\cS})\rangle &=& \nonumber \\ & &\hspace{-30mm}
\frac{\left(\lambda_{\Phi_1\Phi_2\cO}^{(0)}+\lambda_{\Phi_1\Phi_2\cO}^{(1)}z+
\lambda_{\Phi_1\Phi_2\cO}^{(2)}z^{2}\right)S_{-}^{\ell}+\lambda_{\Phi_1\Phi_2\cO}^{(3)}S_{+}S_{-}^{\ell-1}}
{\left(\<1\bar{0}\>\<2\bar{0}\>\right)^{\de} \<1\bar{2}\>^{q_1-\de} \<2\bar{1}\>^{q_2-\de} \<0\bar{2}\>^{(\bar q_2-q_1)+\de} \<0\bar{1}\>^{(\bar q_1-q_2)+\de} }, \label{3p}
\eea
where $\de\equiv\frac{1}{4} (\Delta +\ell -R)$. The numerator contains four free coefficients according to the properties of spin-$\ell$ tensor structures and invariant cross ratio $z$. It is straightforward to show that the denominator satisfies the homogeneity conditions of the three operators, but this is not the
only choice. The homogeneity conditions can only fix the powers of supertraces $\<i\bar j\>$ up to a free parameter. Specifically, one can adjust the
powers of supertraces through the identity
\beq
\left({\frac{\<1\bar 2\>}{\langle 1\bar{0}\rangle\langle 0\bar{2}\rangle}}\right)^{2a}
=
\left({\frac{\langle 1\bar{2}\rangle\langle 2\bar{1}\rangle}{\langle 1\bar{0}\rangle \langle 0\bar{2}\rangle\langle 0\bar{1}\rangle \langle 2\bar{0}\rangle}}\right)^a(1-2a z+2a^2 z^{2}), \label{match}
\eeq
in the meanwhile, the coefficients $\lambda_{\Phi_1\Phi_2\cO}^{(i)}$ will be transformed linearly. In (\ref{3p}) we have adopted a particular
gauge that the supertraces $\<1\bar0\>$ and $\<2\bar0\>$ have identical power. It will be more convenient to compute superconformal integration in this gauge.

\subsection{Remarks on the Complex Coefficients}
For the three-point correlator of scalars with arbitrary superconformal weights, it needs to clarify the relationship between $(\la^{(i)}_{\Phi_1\Phi_2\cO})^*$ and $\la^{(i)}_{\Phi^\dagger_2\Phi^\dagger_1\cO^\dagger}$.

Let us evaluate three-point correlator $\<\Phi^\dagger_2(1,\bar 1)\Phi^\dagger_1(2,\bar 2)\cO^\dagger(0,\bar 0)\>$. We can
directly apply Eq.~(\ref{3p}) with three group of quantum numbers $(0,~0,~\bar q_2, q_2), (0,~0,~\bar q_1, q_1), (\frac{\ell}{2},\frac{\ell}{2}, \De, -R_{\cO})$:
\bea
\<\Phi^\dagger_2(1,\bar 1)\Phi^\dagger_1(2,\bar 2)\cO^\dagger(0,\bar 0)\> &=& \nonumber \\ & &\hspace{-30mm}
\frac{\left(\lambda_{\Phi^\dagger_2\Phi^\dagger_1\cO^\dagger}^{(0)}+\lambda_{\Phi^\dagger_2\Phi^\dagger_1\cO^\dagger}^{(1)}z+
\lambda_{\Phi^\dagger_2\Phi^\dagger_1\cO^\dagger}^{(2)}z^{2}\right)S_{-}^{\ell}+\lambda_{\Phi^\dagger_2\Phi^\dagger_1\cO^\dagger}^{(3)}S_{+}S_{-}^{\ell-1}}
{\left(\<1\bar{0}\>\<2\bar{0}\>\right)^{\de'} \<1\bar{2}\>^{\bar q_2-\de'} \<2\bar{1}\>^{\bar q_1-\de'} \<0\bar{2}\>^{(q_1-\bar q_2)+\de'} \<0\bar{1}\>^{(q_2-\bar q_1)+\de'} }, \label{3pc1}
\eea
where $\de'\equiv\frac{1}{4} (\Delta +\ell +R)$.

Alternatively, we can also solve above three-point correlator by taking Hermitian conjugate on (\ref{3p}) and then permuting coordinates $1\leftrightarrow2$. Both the
invariant cross ratio $z$ and the spin-$\ell$ tensor structure $S$ are invariant under the combination actions of
Hermitian conjugate and coordinate permutation $1\leftrightarrow2$, the new three-point function turns into
\bea
\<\Phi^\dagger_1(2,\bar 2)\Phi^\dagger_2(1,\bar 1)\cO^\dagger(0,\bar 0)\> &=& \nonumber \\ & &\hspace{-30mm}
\frac{\left((\lambda_{\Phi_1\Phi_2\cO}^{(0)})^*+(\lambda_{\Phi_1\Phi_2\cO}^{(1)})^*z+
(\lambda_{\Phi_1\Phi_2\cO}^{(2)})^*z^{2}\right)S_{-}^{\ell}+(\lambda_{\Phi_1\Phi_2\cO}^{(3)})^*S_{+}S_{-}^{\ell-1}}
{\left(\<0\bar{1}\>\<0\bar{2}\>\right)^{\de} \<1\bar{2}\>^{q_1-\de} \<2\bar{1}\>^{q_2-\de} \<1\bar{0}\>^{(\bar q_2-q_1)+\de} \<2\bar{0}\>^{(\bar q_1- q_2)+\de} }. \label{3pc2}
\eea
To compare Eq.~(\ref{3pc2}) with Eq.~(\ref{3pc1}), we need to make a transformation (\ref{match}) in Eq.~(\ref{3pc2}) with parameter
\beq
a=\frac{q_2+\bar q_2-q_1-\bar q_1}{2}=-\frac{r}{2},
\eeq
then the two equations share exactly the same denominator. Identifying the tensor structures in their numerators, we obtain following linear relationships among the complex coefficients
\bea
\lambda_{\Phi^\dagger_2\Phi^\dagger_1\cO^\dagger}^{(0)}&=&(\lambda_{\Phi_1\Phi_2\cO}^{(0)})^*, \nonumber\\
\lambda_{\Phi^\dagger_2\Phi^\dagger_1\cO^\dagger}^{(1)}&=&r(\lambda_{\Phi_1\Phi_2\cO}^{(0)})^*+(\lambda_{\Phi_1\Phi_2\cO}^{(1)})^*, \nonumber \\
\lambda_{\Phi^\dagger_2\Phi^\dagger_1\cO^\dagger}^{(2)}&=&\frac{1}{2}r^2(\lambda_{\Phi_1\Phi_2\cO}^{(0)})^*+
r(\lambda_{\Phi_1\Phi_2\cO}^{(1)})^*+(\lambda_{\Phi_1\Phi_2\cO}^{(2)})^*+\frac{1}{2}r(\lambda_{\Phi_1\Phi_2\cO}^{(3)})^*, \nonumber \\
\lambda_{\Phi^\dagger_2\Phi^\dagger_1\cO^\dagger}^{(3)}&=&(\lambda_{\Phi_1\Phi_2\cO}^{(3)})^*. \label{coefs}
\eea
By taking above complex conjugate transformation of the coefficients twice, we go back to the original coefficients, as expected. The linear transformation turns into trivial $(\la^{(i)}_{\Phi_1\Phi_2\cO})^*=\la^{(i)}_{\Phi^\dagger_2\Phi^\dagger_1\cO^\dagger}$ given $r=0$, i.e., scalars $\Phi_1$ and $\Phi_2$ share the same scaling dimension.

\subsection{Three-point Functions with Chiral Operator}
Three-point function can be significantly simplified if there is a chiral or anti-chiral operator. Results obtained from these short multiplets will provide key elements to compute the most general superconformal blocks.

Let us consider the three-point correlator $\<\Phi(1)X(2,\bar 2)\cO(0,\bar0)\>$ which will be needed to compute the shadow coefficients.
The three-point correlator contains a chiral field $\Phi: (0,~0,~q_1,~0)$, a general field $X:(0,~0,~q_2,~\bar{q}_2)$ and a spin-$\ell$ operator $\cO:(\frac{\ell}{2},~\frac{\ell}{2},~\frac{\De+R}{2},~\frac{\De-R}{2})$, where $R=\bar{q}_2-q_1-q_2$. From the chirality of $\Phi$, we can obtain the simplified three-point function
\bea
\langle \Phi(1)X(2,\bar{2})\mathcal{O}(0,\bar{0},\cS,\bar{\cS})\rangle &=& \nonumber \\
&&\hspace{-40mm}
\frac{\lambda_{\Phi X\mathcal{O}} ~S^{\ell}}{\langle 1\bar{2}\rangle ^{\frac{1}{2}(q_1+q_2+\bar{q}_2-\Delta-\ell)} \langle 1\bar{0}\rangle^{\frac{1}{2}(q_1-q_2-\bar{q}_2+\Delta+\ell)} \langle 2\bar{0}\rangle^{q_2}
\langle 0\bar{2}\rangle^{\frac{1}{2}(-q_1-q_2+\bar{q}_2+\Delta+\ell)}}. \label{pxo}
\eea
Taking the transformation (\ref{match}) with $a=\frac{1}{4}(\Delta+\ell+2r+R)$, where $r=q_1-q_2-\bar{q}_2$ in this problem, above equation turns into
\bea
\langle \Phi(1)X(2,\bar{2})\mathcal{O}(0,\bar{0},\cS,\bar{\cS})\rangle &=& \nonumber \\
&&\hspace{-35mm}
=\lambda_{\Phi X\mathcal{O}}\frac{
(1-2a z +a\,(2a-\ell)z^2)S_{-}^{\ell}+\ell S_{+}S_{-}^{\ell-1}}{\langle 1\bar{2}\rangle ^{q_1-q_2-a} \langle 2\bar{1}\rangle^{-a}
(\langle 1\bar{0}\rangle\langle 2\bar{0}\rangle)^{a+q_2} \langle 0\bar{1}\rangle^{a}\langle 0\bar{2}\rangle^{a+q_2+\bar{q}_2-q_1}}, \label{chiX}
\eea
which is consistent with the general three-point function (\ref{3p}) given $\bar q_1=0,~\de=a+q_2$. The
four free coefficients are fixed by the chirality condition up to an overall constant. Such kind of three-point function with real $X$ appears in bootstrapping
the mixed correlator of minimal $4D$ $N=1$ SCFT. In the theory the scalar $X$ appears in OPE $\Phi\times\Phi^\dagger$ so is real: $q_2=\bar{q}_2$.

Similarly, one can use anti-chirality condition to partially fix the coefficients in three-point function $\<\Phi(\bar{1})^\dagger X(2,\bar 2)\cO(0,\bar0)\>$:
\beq
(\la_{\Phi^\dagger X\cO}^{(0)},~ \la_{\Phi^\dagger X\cO}^{(2)}, ~\la_{\Phi^\dagger X\cO}^{(1)},~ \la_{\Phi^\dagger X\cO}^{(3)})=\la_{\Phi^\dagger X\cO}(1,\, a'(2a'-\ell),\,-2a',\, \ell),
\eeq
where $a'=\frac{1}{4}(\De+\ell-R)$, $R=\bar{q_1}+\bar{q}_2-q_2$.

\section{Superconformal Partial Waves}
Now we are ready to study the most general four-point correlator
\beq
\<\Phi_1(1,\bar1)\Phi_2(2,\bar2)\Phi_3(3,\bar3)\Phi_4(4,\bar4)\>,
\eeq
where $\Phi_i$ have arbitrary superconformal weights $(q_i,\bar q_i)$ constrained by vanishing net R-charges
\beq
\sum_iq_i-\sum_i\bar q_i=0.
\eeq
Here we are interested in the superconformal partial wave which gives the amplitude of exchanging an irreducible representation of the $\mathcal{N}=1$ superconformal group. Let us denote such irreducible representation by its superprimary field $\cO:(\frac{\ell}{2},\frac{\ell}{2},\De,R_{\cO})$.
By inserting the projector constructed from $\cO$ and its shadow operator $\tl\cO$ into the four-point correlator, the superconformal partial wave $\cW_\cO$
becomes
\bea
\mathcal{W}_{\mathcal{O}}&\propto&\langle \Phi_{1}\Phi_{2} \left|\mathcal{O}\right|\Phi_{3}\Phi_{4}\rangle=\int D[0,\bar{0}]\langle \Phi_{1}\Phi_{2} \mathcal{O} (0,\bar{0},\cS,\bar{\cS})\rangle \overleftrightarrow{\mathcal{D}_{\ell}} \langle \tl{\mathcal{O}} (0,\bar{0},\cT,\bar{\cT})\Phi_{3}\Phi_{4} \rangle \nonumber\\
&=&\frac{1}{\<1\bar{2}\>^{q_1-\delta } \<2\bar{1}\>^{q_2-\delta } \<3\bar{4}\>^{q_3-\delta '} \<4 \bar{3}\>^{q_4-\delta '}} \times  \\
&& \hspace{0mm} \int D[0,\bar{0}]\frac{\mathcal{N}_{\ell}^{f}}
{(\<1\bar{0}\> \<2\bar{0}\>)^{\delta } (\<3 \bar{0}\> \<4\bar{0}\>)^{\delta'} \<0\bar{2}\>^{\delta +\bar q_2-q_1}  \<0\bar{1}\>^{\delta +\bar q_1-q_2} \<0\bar{4}\>^{\delta '+\bar q_4-q_3} \<0\bar{3}\>^{\delta '+\bar q_3-q_4}}, \nonumber \label{pwave}
\eea
where $\de=\frac{\De+\ell-R}{4} $, $\de'=\frac{2+R+\ell-\De}{4}$ and $\overleftrightarrow{\mathcal{D}_{\ell}}\equiv\frac{1}{\ell!^{4}}(\partial_{\cS}0\partial_{\cT})^{\ell}
(\partial_{\bar{\cS}}\bar{0}\partial_{\bar{\cT}})^{\ell}$.
$\mathcal{N}_\ell^{f}$  represents the tensor structures as defined in \cite{Khandker:2014mpa}:
\bea
\mathcal{N}_{\ell}^{f}&=&
\left((\lambda^{(0)}_{\Phi_1\Phi_2\cO}+\lambda^{(1)}_{\Phi_1\Phi_2\cO}z+\lambda^{(2)}_{\Phi_1\Phi_2\cO}z^{2})S_{-}^{\ell}
+\lambda^{(3)}_{\Phi_1\Phi_2\cO}S_{+}S_{-}^{\ell-1}\right) \nonumber\\
&& \overleftrightarrow{\mathcal{D}_{\ell}}
\left((\lambda^{(0)}_{\Phi_3\Phi_4\tl{\cO}}+\lambda^{(1)}_{\Phi_3\Phi_4\tl{\cO}}\tl{z}+
\lambda^{(2)}_{\Phi_3\Phi_4\tl{\cO}}\tl{z}^{2})T_{-}^{\ell}+\lambda^{(3)}_{\Phi_3\Phi_4\tl{\cO}}T_{+}T_{-}^{\ell-1}\right),\label{Nf}
\eea
In (\ref{pwave}) we have applied the three-point function
\bea
\langle \Phi_{3}(3,\bar{3})\Phi_{4}(4,\bar{4})\tl\cO(0,\bar{0},\cT,\bar{\cT})\rangle &=& \nonumber \\ & &\hspace{-30mm}
\frac{\left(\lambda_{\Phi_3\Phi_4\tl\cO}^{(0)}+\lambda_{\Phi_3\Phi_4\tl\cO}^{(1)}\tl{z}+
\lambda_{\Phi_3\Phi_4\tl\cO}^{(2)}\tl{z}^{2}\right)T_{-}^{\ell}+\lambda_{\Phi_3\Phi_4\tl\cO}^{(3)}T_{+}T_{-}^{\ell-1}}
{\left(\<3\bar{0}\>\<4\bar{0}\>\right)^{\de'} \<3\bar{4}\>^{q_3-\de'} \<4\bar{3}\>^{q_4-\de'} \<0\bar{4}\>^{(\bar q_4-q_3)+\de'} \<0\bar{3}\>^{(\bar q_3-q_4)+\de'} }. \label{3ps}
\eea
where $(\tl{z},~T_{\pm}^\ell)$, like $(z,~S_{\pm}^\ell)$ in (\ref{3p}), are invariant cross ratio and tensor structures. Tensor structures
in $\cN_\ell^f$ consist of the polynomial $\cN_\ell$
\beq
\cN_{\ell}\equiv(\bar{\cS}1\bar{2}\cS)^{\ell}\overleftrightarrow{\mathcal{D}_{\ell}}(\bar{\cT}3\bar{4}\cT)^{\ell} 
\eeq
 and its coordinate exchanges.
Giving $\theta_{\rm ext}=\bar{\theta}_{\rm ext}=0$ and $\cX_0=\bar\cX_0$, $\cN_{\ell}$ reduces to
\begin{equation}
N_{\ell} = y_0^{\frac{\ell}{2}} C_{\ell}^{(1)}(y_{0}),
\end{equation}
where $C_{\ell}^{(\la)}(y )$ are the Gegenbauer polynomials and
\bea
x_{0}&\equiv &-\frac{X_{13}X_{20}X_{40}}{2\sqrt{X_{10}X_{20}X_{30}X_{40}X_{12}X_{34}}}-\left(1\leftrightarrow2\right)-\left(3\leftrightarrow4\right), \\
y_{0}&\equiv &\frac{1}{2^{12}}X_{10}X_{20}X_{30}X_{40}X_{12}X_{34}.
\eea

For the four-point function of scalars we are only interested in the lowest component of a supermultiplet. To throw away irrelevant
higher dimensional components we set the fermionic coordinates $\theta_{\rm ext}=\bar{\theta}_{\rm ext}=0$. The  bi-supertwistors
$\cX_{AB},\tl\cX^{AB}$ degenerate into twistors $X_{\alpha\beta}, X^{\alpha\beta}$ which are equivalent to the six dimensional vector representations of $\rm SU(2,2)\cong\rm SO(4,2)$, and the supertraces $\<i\bar j\>$ become inner products of vectors $ X_{ij} \equiv -2X_i \cdot X_j$.
Moreover, under the restriction $\theta_{\rm ext}=\bar{\theta}_{\rm ext}=0$ the superconformal integration (\ref{pwave}) can be simplified into nonsupersymmetric conformal integration,
as suggested in (\ref{Int1}). To summarize, the superconformal partial wave $\cW_\cO$ is
\begin{equation}
\mathcal{\left.W_{O}\right|}_{\theta_{\mathrm{ext}}=0}\propto\frac{1}{X_{12}^{q_1+q_2-2\de}X_{34}^{q_3+q_4-2\de'}}\int D^{4}X_{0}\left.\partial_{\bar{0}}^{2}\frac{\cN_{\ell}^{f}}{D_{\ell}}\right|_{\bar{0}=0},
\label{pwave1}
\end{equation}
and $D_{\ell}$ denotes the products of supertraces containing $\cX_0$ or $\bar\cX_0$
\begin{equation}
 D_{\ell}\equiv (X_{1\bar{0}}X_{2\bar{0}})^{\delta } (X_{3 \bar{0}} X_{4\bar{0}})^{\delta'} X_{0\bar{2}}^{\delta +\bar q_2-q_1}  X_{0\bar{1}}^{\delta +\bar q_1-q_2} X_{0\bar{4}}^{\delta '+\bar q_4-q_3} X_{0\bar{3}}^{\delta '+\bar q_3-q_4}. \label{Dl}
\end{equation}
As shown in (\ref{pwave1}), essentially there are only two steps to accomplish the superconformal integration for $\cW_\cO$:
partial derivatives on $ \cN_{\ell}^{f}/D_{\ell}$ and conformal integration. The partial derivatives are straightforward to evaluate. The conformal integration related to Gegenbauer polynomial $C_{\ell}^{(1)}(x_{0})$ has been detailedly studied in \cite{Dolan:2000ut, SimmonsDuffin:2012uy}. Since the result is fundamental for our study we repreat it here for convenience
{\small
\begin{eqnarray}
\int_M D^{4}X_{0}\frac{(-1)^{\ell} C_{\ell}^{(1)}(x_{0})}{X_{10}^{\frac{\Delta+r}{2}}X_{20}^{\frac{\Delta-r}{2}}X_{30}^{\frac{\tl{\Delta}+\tl r}{2}}X_{40}^{\frac{\tl{\Delta}-\tl r}{2}}}=\xi_{\Delta,\tl{\Delta},\tl r,\ell}\left(\frac{X_{14}}{X_{13}}\right)^{\frac{\tl r}{2}}\left(\frac{X_{24}}{X_{14}}\right)^{\frac{r}{2}}X_{12}^{-\frac{\Delta}{2}}X_{34}^{-\frac{\tl{\Delta}}{2}}g_{\Delta,\ell}^{r,\tl r}(u,v),
\label{Cint}
\end{eqnarray}
}
in which $r\equiv\De_1-\De_2$, $\tl r\equiv \De_3-\De_4$ and
\begin{equation}
\xi_{\Delta,\tl{\Delta},\tl r,\ell}\equiv\frac{\pi^{2}\Gamma(\tl{\Delta}+\ell-1)
\Gamma(\frac{\Delta-\tl r+\ell}{2})\Gamma(\frac{\Delta+\tl r+\ell}{2})}{(2-\Delta)\Gamma(\Delta+\ell)
\Gamma(\frac{\tl{\Delta}-\tl r+\ell}{2})\Gamma(\frac{\tl{\Delta}+\tl r+\ell}{2})}.\label{xi}
\end{equation}
The conformal blocks $g_{\Delta,\ell}^{r,\tl r}(u,v)$ are defined as usual
\bea
g_{\Delta,\ell}^{r,\tl r}(u,v)&=&\frac{\rho\bar{\rho}}{\rho-\bar{\rho}}\left[k_{\Delta+\ell}(\rho)k_{\Delta-\ell-2}(\bar{\rho})-(\rho\leftrightarrow\bar{\rho})\right]\label{g}, \nonumber\\
k_{\beta}(x)&=&x^{\frac{\beta}{2}}{}_{2}F_{1}\left(\frac{\beta-r}{2},\frac{\beta+\tl r}{2},\beta,x\right),
\eea
where $u,\,v$ are the standard conformal invariants and $u=\rho\bar{\rho},~v=(1-\rho)(1-\bar{\rho})$.

To apply above results on conformal integrations in our case, the most crucial step is to
write the integrand into a compact form in terms of Gegenbauer polynomials.

Giving $\theta_{\rm ext}=\bthex=0$, the only non-vanishing
fermionic coordinates are $\theta_0,\bar\theta_0$ from bisupertwistors $\cX_0, \bar\cX_0$. Supconformal invariants proportional to
the fermionic coordinates therefore vanish at third or higher orders. Moreover, as shown in \cite{Khandker:2014mpa},
 the tensor structure terms in $\cN_\ell^f$ can be separated into symmetric
$(\cN_\ell^+)$ or antisymmetric $(\cN_\ell^-)$ parts according to their performances under coordinate interchange $1\leftrightarrow3,~2\leftrightarrow4$:
\beq
\cN_\ell^f=\cN_\ell^++\cN_\ell^-,
\eeq
in which
\bea
\cN_{\ell}^{+}&=&S_{-}^{\ell}\overleftrightarrow{\mathcal{D}_{\ell}}T_{-}^{\ell}\left(\lambda^{(0)}_{\Phi_1\Phi_2\cO}\lambda^{(0)}_{\Phi_3\Phi_4\tl{\cO}} +\lambda^{(2)}_{\Phi_1\Phi_2\cO}\lambda^{(0)}_{\Phi_3\Phi_4\tl{\cO}}z^{2}
+\lambda^{(0)}_{\Phi_1\Phi_2\cO}\lambda^{(2)}_{\Phi_3\Phi_4\tl{\cO}}\tl{z}^{2} \right.  \nonumber \\
&&\hspace{18mm}\left.+\lambda^{(1)}_{\Phi_1\Phi_2\cO}\lambda^{(1)}_{\Phi_3\Phi_4\tl{\cO}}z\tl{z} \right)
+S_{-}^{\ell}\overleftrightarrow{\mathcal{D}_{\ell}}T_{+}T_{-}^{\ell-1} \lambda^{(1)}_{\Phi_1\Phi_2\cO}\lambda^{(3)}_{\Phi_3\Phi_4\tl{\cO}}z  \nonumber \\
&&+S_{+}S_{-}^{\ell-1}\overleftrightarrow{\mathcal{D}_{\ell}}T_{-}^{\ell} \lambda^{(3)}_{\Phi_1\Phi_2\cO}\lambda^{(1)}_{\Phi_3\Phi_4\tl{\cO}}\tl{z}
+S_{+}S_{-}^{\ell-1}\overleftrightarrow{\mathcal{D}_{\ell}}T_{+}T_{-}^{\ell-1} \lambda^{(3)}_{\Phi_1\Phi_2\cO}\lambda^{(3)}_{\Phi_3\Phi_4\tl{\cO}}, \label{NF2}
\eea
and
\bea
\cN_{\ell}^{-}&=& zS_{-}^{\ell}\overleftrightarrow{\mathcal{D}_{\ell}}T_{-}^{\ell}\lambda^{(1)}_{\Phi_1\Phi_2\cO}\lambda^{(0)}_{\Phi_3\Phi_4\tl{\cO}}
+\tl{z}S_{-}^{\ell}\overleftrightarrow{\mathcal{D}_{\ell}}
T_{-}^{\ell}\lambda^{(0)}_{\Phi_1\Phi_2\cO}\lambda^{(1)}_{\Phi_3\Phi_4\tl{\cO}} \nonumber\\
&&+ S_{-}^{\ell}\overleftrightarrow{\mathcal{D}_{\ell}}T_{+}T_{-}^{\ell-1}
\lambda^{(0)}_{\Phi_1\Phi_2\cO}\lambda^{(3)}_{\Phi_3\Phi_4\tl{\cO}} + S_{+}S_{-}^{\ell-1}\overleftrightarrow{\mathcal{D}_{\ell}}T_{-}^{\ell}
\lambda^{(3)}_{\Phi_1\Phi_2\cO}\lambda^{(0)}_{\Phi_3\Phi_4\tl{\cO}} .\label{anti}
\eea

Contributions of the symmetric terms $\cN_\ell^+$ on the superconformal partial wave $\cW_\cO$ have been detailedly studied in \cite{Khandker:2014mpa} under the restrictions
\beq
q_1=\bar q_2, ~q_2=\bar q_1, ~q_3=\bar q_4, ~q_4=\bar q_3. \label{limt}
 \eeq
Under above restrictions the coordinate interchange symmetry in $\cN^+_\ell$ is further realized in the whole integrand of superconformal partial wave $\cW_\cO$, and due
 to this symmetry, it gets much simpler to evaluate contributions on $\cW_\cO$ from the symmetric terms.
While for the most general superconformal partial waves we do not have such restrictions on
the superconformal weights, nevertheless, there is a free parameter related to the transformation (\ref{match}), and we can choose the gauge in which $X_{1\bar0} ~(X_{3\bar0})$ and $X_{2\bar 0} ~(X_{4\bar0})$ have the same power, then it is straightforward to calculate contributions of these terms on $\cW_\cO$. More details on the calculations are provided in Appendix B.

The major challenge comes from the four terms in $\cN_\ell^-$ which are anti-symmetric under the coordinate interchange $1\leftrightarrow3,~2\leftrightarrow4$ (anti-symmetric terms).
For the cases studied in \cite{Khandker:2014mpa}, due to the restrictions (\ref{limt}), $D_\ell$ is invariant under coordinate interchange $1\leftrightarrow3,~2\leftrightarrow4$, and contributions from anti-symmetric terms are cancelled automatically. While for general four-point functions there is no such
coordinate interchange symmetry in $D_\ell$, and contributions from terms in (\ref{anti}) are proportional to the differences of scaling dimensions $r,~\tl r$.

\subsection{Superconformal Integrations of Anti-symmetric Terms}
In this section we evaluate superconformal integrations of the anti-symmetric terms in (\ref{anti}) following the strategy discussed before.
However, to apply the conformal integration formulas in (\ref{Cint}), we need to figure out relationships between tensor structures in $\cN_\ell^-$
and the Gegenbauer polynomials. For tensor structures in $\cN_\ell^+$, the polynomials satisfy coordinate interchange symmetry and can be simplified
 using Clifford algebra. Nevertheless, for tensor structures in $\cN_\ell^-$, the polynomials are anti-symmetric under coordinate permutation and
the Clifford algebra cannot help to simplify the polynomials directly, instead, we show that these polynomials possesses recursion relations which can 
be used to determine the superconformal integrations.

The anti-symmetric terms in (\ref{anti}) consist of $\frac{z N_\ell}{D_\ell}$, $\frac{\tl z N_\ell}{D_\ell}$, $\frac{N_\ell}{D_\ell}$ and their coordinate exchanges. The partial differentiations are
{\small 
\begin{eqnarray}
\partial_{\bar{0}}^{2}\frac{zN_{\ell}}{D_{\ell}}|_{\bar{0}=0}&=&2\delta'\frac{N_{\ell}}{D_{\ell}}\left[\frac{X_{13}}{X_{10}X_{30}}
-\frac{X_{23}}{X_{20}X_{30}}+\frac{X_{14}}{X_{10}X_{40}}-\frac{X_{24}}{X_{20}X_{40}}\right] \nonumber \\
&&  +\frac{1}{2}\frac{1}{D_{\ell}}\frac{\ell}{\ell!^{2}}\left(\partial_{S}0\partial_{T}\right)^{\ell}\left(S\bar{2}1\bar{0}3\bar{4}T\right)^{\ell-1}\left[\frac{X_{12}}{X_{10}X_{20}}X_{10}\left(S\bar{2}3\bar{4}T\right)\right], \\
\left.\partial_{\bar{0}}^{2}\frac{\tl{z}N_{\ell}}{D_{\ell}}\right|_{\bar{0}=0}&=&2\delta\frac{N_{\ell}}{D_{\ell}}\left[\frac{X_{13}}{X_{10}X_{30}}+
\frac{X_{23}}{X_{20}X_{30}}-\frac{X_{14}}{X_{10}X_{40}}-\frac{X_{24}}{X_{20}X_{40}}\right] \nonumber \\
&&  +\frac{1}{2}\frac{1}{D_{\ell}}\frac{\ell}{\ell!^{2}}\left(\partial_{S}0\partial_{T}\right)^{\ell}\left(S\bar{2}1\bar{0}3\bar{4}T\right)^{\ell-1}\left[\frac{X_{34}}{X_{30}X_{40}}X_{30}\left(S\bar{2}1\bar{4}T\right)\right], \\
\left.\partial_{\bar{0}}^{2}\frac{N_{\ell}}{D_{\ell}}\right|_{\bar{0}=0}&=&-\frac{N_{\ell}}{D_{\ell}}\left[4 \delta^{2}\frac{X_{12}}{X_{10}X_{20}}
+4\delta'^{2}\frac{X_{34}}{X_{30}X_{40}} +4 \delta\delta'\left(\frac{X_{13}}{X_{10}X_{30}}
+\frac{X_{23}}{X_{20}X_{30}}+\frac{X_{14}}{X_{10}X_{40}}\right.\right. \nonumber \\
&& \left.\left.+\frac{X_{24}}{X_{20}X_{40}}\right)\right]+\frac{1}{2}\frac{1}{D_{\ell}}\frac{\ell}{\ell!^{2}}\left(\partial_{S}0\partial_{T}\right)^{\ell}\left(S\bar{2}1\bar{0}3\bar{4}T\right)^{\ell-1}
\left[2\delta\frac{X_{12}}{X_{10}X_{20}}\left(X_{10}S\bar{2}3\bar{4}T\right) \right. \nonumber \\
&&
\left. +2\delta'\frac{X_{34}}{X_{30}X_{40}}\left(X_{30}S\bar{2}1\bar{4}T\right)\right] .
\end{eqnarray}}
For the terms proportional to $N_\ell$, their conformal integrations can be evaluated directly using Eq.~(\ref{Cint}), the results are provided in Appendix B.
While for extra terms, we need to find their relationships with Gegenbauer polynomials before we can apply Eq.~(\ref{Cint}).
Tensor structures in (\ref{anti}) can be expanded in terms of $\cN_\ell$ and its coordinate exchanges as
\begin{eqnarray}
S_{-}^{\ell}\overleftrightarrow{\mathcal{D}_{\ell}}T_{-}^{\ell}&=&\frac{\cN_{\ell}}{4\langle 1\bar{2}\rangle ^{\ell}\langle 3\bar{4}\rangle ^{\ell}}+(-1)^{\ell}(1\leftrightarrow2)+(-1)^{\ell}(3\leftrightarrow4),  \\
S_{-}^{\ell}\overleftrightarrow{\mathcal{D}_{\ell}}T_{+}T_{-}^{\ell-1}&=&\frac{\cN_{\ell}}{4\ell\langle 1\bar{2}\rangle ^{\ell}\langle 3\bar{4}\rangle ^{\ell}}+(-1)^{\ell}(1\leftrightarrow2)-(-1)^{\ell}(3\leftrightarrow4),  \\
S_{+}S_{-}^{\ell-1}\overleftrightarrow{\mathcal{D}_{\ell}}T_{-}^{\ell}&=&\frac{\cN_{\ell}}{4\ell\langle 1\bar{2}\rangle ^{\ell}\langle 3\bar{4}\rangle ^{\ell}}-(-1)^{\ell}(1\leftrightarrow2)+(-1)^{\ell}(3\leftrightarrow4),
\label{Symm}
\end{eqnarray}%
which lead to following polynomial terms in the conformal integrand
\bea
R_\ell\equiv \frac{\ell}{{\ell!}^2} \left(\partial_{S}0\partial_{T}\right)^{\ell}\left(S\bar{2}1\bar{0}3\bar{4}T\right)^{\ell-1}\times&&\nonumber \\
 && \hspace{-40mm}  \left( X_{10}S\bar{2}3\bar{4}T + X_{20}S\bar{1}3\bar{4}T - X_{10}S\bar{2}4\bar{3}T - X_{20}S\bar{1}4\bar{3}T\right),  \label{R0}\\
P_\ell\equiv \frac{\ell}{{\ell!}^2}  \left(\partial_{S}0\partial_{T}\right)^{\ell} \left(S\bar{2}1\bar{0}3\bar{4}T\right)^{\ell-1} \times&& \nonumber\\
&& \hspace{-40mm} \left( X_{30}S\bar{2}1\bar{4}T + X_{40}S\bar{2}1\bar{3}T - X_{30}S\bar{1}2\bar{4}T - X_{40}S\bar{1}2\bar{3}T\right).  \label{P0}
\eea
It is shown in Appendix A that above polynomials satisfy the recursion relations
\bea
R_\ell &=& \ell\De_A N_{\ell-1}+\frac{1}{2^6}(\ell-1)X_{10}X_{20}X_{34}\De_B N_{\ell-2}+y_0R_{\ell-2}, \label{Rlr0}\\
P_\ell &=& \ell\De_B N_{\ell-1}+\frac{1}{2^6}(\ell-1)X_{30}X_{40}X_{12}\De_A N_{\ell-2}+y_0P_{\ell-2}. \label{Plr0}
\eea
%
The conformal integrations related to $R_\ell$ and $P_\ell$ are
%
{\small
\begin{eqnarray}
\hspace{-5mm} \int D^{4}X_{0}\left. \frac{X_{12}}{X_{10}X_{20}}\frac{R_\ell}{D_\ell}\right|_{\bar{0}=0}
&=&\frac{8\,c_{\ell}\,\xi_{\De+2,2-\De,1+\tl{r},\ell-1}}{X_{12}^{\frac12(\De-\ell)}X_{34}^{-\frac12(\De+\ell-2)}}
\left(\frac{X_{24}}{X_{14}}\right)^{\frac{r}{2}}\left(\frac{X_{14}}{X_{13}}\right)^{\frac{\tl{r}}{2}}\times \nonumber\\
&&\left[
-\frac{4  \tl r \Delta (\ell +1) (\Delta -\ell )}{(\Delta -1) (\Delta +\tl r-\ell ) (\Delta +\tl r+\ell )}
 g_{\De+1,\ell-1}^{r,\tl{r}} \right.\nonumber\\
&&
\left.+\frac{r \ell  (\Delta -\ell ) (\Delta -\tl r+\ell )}{(\Delta +\ell ) (\Delta +\ell +1) (\Delta +\tl r-\ell )}
g_{\De+2,\ell}^{r,\tl{r}} \right],  \label{anti1}  \\
\hspace{-5mm} \int D^{4}X_{0}\left. \frac{X_{34}}{X_{30}X_{40}}\frac{P_\ell}{D_\ell}\right|_{\bar{0}=0}
&=&\frac{8\,c_{\ell}\,\xi_{\De,4-\De,1+\tl{r},\ell-1}}{X_{12}^{\frac12(\De-\ell)}X_{34}^{-\frac12(\De+\ell-2)}}
\left(\frac{X_{24}}{X_{14}}\right)^{\frac{r}{2}}\left(\frac{X_{14}}{X_{13}}\right)^{\frac{\tl{r}}{2}}\times \nonumber\\
&&\left[
-\frac{r(\Delta -2) (\ell +1) (-\Delta +\tl r+\ell +2) (\Delta -\tl r+\ell -2)}{4 (\Delta -1)
(-\Delta +\ell +1) (-\Delta +\ell +2) (\Delta +\ell -1)}
 g_{\De+1,\ell-1}^{r,\tl{r}} \right.\nonumber\\
&&
\left.-\frac{\tl r \ell  (\Delta -\tl r+\ell -2)}{(\Delta +\ell -1) (\Delta +\tl r-\ell -2)}
g_{\De,\ell}^{r,\tl{r}} \right], \label{anti2}
\end{eqnarray}}
where $c_\ell=2^{-6\ell}$.
Above equations can be proved using mathematical induction based on the recursion relations (\ref{Rlr0}) and (\ref{Plr0}).
Conformal integrations in (\ref{anti1}) and (\ref{anti2}), together with the results presented in Appendix B, provide all the necessary materials to
compute the superconformal partial waves $\cW_\cO$ for general scalars $\Phi_i$. Here we present the final results of superconformal partial wave
(\ref{pwave1}):
\bea
\cW_\cO&\propto& \frac{1}{X_{12}^{\frac{\De_1+\De_2}{2}}X_{34}^{\frac{\De_3+\De_4}{2}}}\left(\frac{X_{24}}{X_{14}}\right)^{\frac{r}{2}}
\left(\frac{X_{14}}{X_{13}}\right)^{\frac{\tl{r}}{2}}\times \nonumber\\
&&\hspace{30mm}\left(a_1\, g_{\De,\ell}^{r,\tl{r}}+ a_2\, g_{\De+1,\ell+1}^{r,\tl{r}}+ a_3\, g_{\De+1,\ell-1}^{r,\tl{r}}
+ a_4\, g_{\De+2,\ell}^{r,\tl{r}}\right), \label{pwavef}
\eea
in which the coefficients $a_i$ are the abbreviations of following long expressions:
{
\bea
a_1&=&2 \la_{\Phi_1\Phi_2\cO}^{(0)} \left[ -\de' \left(1+2 \delta\frac{(2-\Delta ) \tl r^2-(\ell +2-\Delta) (\Delta +\ell )}{(\Delta -1) ( \ell +2-\Delta) (\Delta +\ell )}\right)\la_{\Phi_3\Phi_4\tl\cO}^{(0)}+\la_{\Phi_3\Phi_4\tl\cO}^{(2)} \right.\nonumber \\
&& \left.+\frac{\tl r ((\Delta -2) R+(-\Delta +\ell +2) (\Delta +\ell ))}{2 (-\Delta +\ell +2) (\Delta +\ell )}\la_{\Phi_3\Phi_4\tl\cO}^{(1)}+\frac{\tl r (R+\ell +2-\Delta)}{4 (-\Delta +\ell +2)}\la_{\Phi_3\Phi_4\tl\cO}^{(3)}
\right],\label{a1} \\
a_2&=&- \frac{(\Delta -2) (\Delta -\tl r+\ell ) (\Delta +\tl r+\ell )}{4 (\Delta -1) (\Delta +\ell ) (\Delta +\ell +1)} \nonumber \\
&&\times\left(\la_{\Phi_1\Phi_2\cO}^{(1)}+  \frac{r (\Delta -R+\ell )}{2 (\Delta +\ell )}\la_{\Phi_1\Phi_2\cO}^{(0)} \right)
\left(\la_{\Phi_3\Phi_4\tl\cO}^{(1)} +\frac{\tl r (R+\ell +2-\Delta )}{2 (-\Delta +\ell +2)} \la_{\Phi_3\Phi_4\tl\cO}^{(0)}\right), \label{a2} \\
a_3&=&-\frac{(\Delta -2) (\Delta -\tl r-\ell -2) (\Delta +\tl r-\ell -2)}{4 (\Delta -1) (-\Delta +\ell +1) (-\Delta +\ell +2)}  \nonumber \\
&&\times\left(\la_{\Phi_1\Phi_2\cO}^{(1)}+ \frac{\ell+1}{\ell}\la_{\Phi_1\Phi_2\cO}^{(3)}+\frac{r (-\Delta +R+\ell +2)}{2 (-\Delta +\ell +2)}\la_{\Phi_1\Phi_2\cO}^{(0)} \right) \nonumber\\
&&\times\left(\la_{\Phi_3\Phi_4\tl\cO}^{(1)} +\frac{\ell+1}{\ell}\la_{\Phi_3\Phi_4\tl\cO}^{(3)}+\frac{\tl r (\Delta -R+\ell )}{2 (\Delta +\ell )}\la_{\Phi_3\Phi_4\tl\cO}^{(0)}\right),  \label{a3}\\
a_4 &=& 2\la_{\Phi_3\Phi_4\tl\cO}^{(0)}\frac{(\Delta -2) (-\Delta -\tl r+\ell +2) (-\Delta +\tl r+\ell +2) (\Delta -\tl r+\ell ) (\Delta +\tl r+\ell )}{16 \Delta  (-\Delta +\ell +1) (-\Delta +\ell +2) (\Delta +\ell ) (\Delta +\ell +1)} \nonumber\\
&&\times\left[ -\delta  \left(1-2 \de'\frac{ \left(r^2 \Delta  -(\Delta +\ell ) (-\Delta +\ell +2)\right)}{(\Delta -1) (-\Delta +\ell +2) (\Delta +\ell )}\right)
\la_{\Phi_1\Phi_2\cO}^{(0)}+ \la_{\Phi_1\Phi_2\cO}^{(2)}\right. \nonumber \\
&&\left.+\frac{r (\Delta  (-\Delta +R+2)+\ell  (\ell +2))}{2 (-\Delta +\ell +2) (\Delta +\ell )}\la_{\Phi_1\Phi_2\cO}^{(1)}+\frac{r (\Delta -R+\ell )}{4 (\Delta +\ell )}\la_{\Phi_1\Phi_2\cO}^{(1)}\right]. \label{a4}
\eea}
Several interesting properties appear in above long expressions of coefficients $a_i$. Ignoring the constant term, $a_1$ and
$a_4$ are related to each other through a transformation
\beq
\De\leftrightarrow2-\De,~~r\leftrightarrow \tl r,~~R\leftrightarrow-R, ~~\la_{\Phi_1\Phi_2\cO}^{(i)}\leftrightarrow \la_{\Phi_3\Phi_4\tl\cO}^{(i)},
\eeq
while $a_2$ and $a_3$ are invariant under this transformation. Such symmetry is expected since it corresponds to exchange the roles of operator $\cO$ and its supershadow operator $\tl\cO$.

\section{Superconformal Blocks}
Conformal blocks are obtained from conformal partial waves by dropping some less interesting factors. The $\cN=1$ superconformal block $\cG_{\De,\ell}^{r,\tl r}$ is related to the superconformal partial wave $\cW_\cO$ through
\beq
\cG_{\De,\ell}^{r,\tl r}=X_{12}^{\frac{\De_1+\De_2}{2}}X_{34}^{\frac{\De_3+\De_4}{2}}\left(\frac{X_{24}}{X_{14}}\right)^{-\frac{r}{2}}
\left(\frac{X_{14}}{X_{13}}\right)^{-\frac{\tl{r}}{2}}\cW_\cO.
\eeq
Then applying the results on $\cW_\cO$ (\ref{pwavef}-\ref{a4}) one can get the superconformal block in terms of $\la_{\Phi_1\Phi_2\cO}^{(i)}$ and
$\la_{\Phi_3\Phi_4\tl\cO}^{(i)}$. The supershadow coefficients $\la_{\Phi_3\Phi_4\tl\cO}^{(i)}$ need to be transformed into the normal coefficients $\la_{\Phi_3\Phi_4\cO^\dagger}^{(i)}$. In principle, one can solve the transformation between the two types of coefficients by inserting the integral expression of supershadow
operator $\tl\cO$  (\ref{shadow})
in the three-point function $\langle \Phi_{3}(3,\bar{3})\Phi_{4}(4,\bar{4})\tl\cO(0,\bar{0},\cT,\bar{\cT})\rangle$ (\ref{3ps}). However it needs to
evaluate a complex superconformal integration to obtain the results. A simpler method is proposed in \cite{Khandker:2014mpa} which applies the unitarity of SCFTs. In this work the unitarity of SCFTs is also employed to solve the transformation of supershadow coefficients.

Giving $\Phi_3=\Phi_2^\dagger$ and $\Phi_4=\Phi_1^\dagger$, unitarity of the four-point function $\<\Phi_1\Phi_2\Phi_2^\dagger\Phi_1^\dagger\>$ requires
the coefficients $a_i$ (\ref{a1}-\ref{a4}) of four conformal blocks in $\cG_{\De,\ell}^{r,\tl r}$ to be positive. To apply the unitary condition we need to go back to the coefficients $(\la_{\Phi_1\Phi_2\cO}^{(i)})^*$   rather than use $\la_{\Phi_2^\dagger\Phi_1^\dagger\cO^\dagger}^{(i)}$ directly.
At first it is not clear whether there is a linear map connecting $\la_{\Phi_2^\dagger\Phi_1^\dagger\tl\cO}^{(i)}$ with
$(\la_{\Phi_1\Phi_2\cO}^{(i)})^*$. Possible
transformations among the three types of coefficients are shown in graph as below

\begin{displaymath}
\xymatrixcolsep{5pc}\xymatrix{
\la_{\Phi_2^\dagger\Phi_1^\dagger\tl\cO}^{(i)} ~~ \ar[dr]_{H_0}  \ar[r]^{H_1} & ~~~(\la_{\Phi_1\Phi_2\cO}^{(i)})^* \ar[d]^{H_2}\\
  & ~~\la_{\Phi_2^\dagger\Phi_1^\dagger\cO^\dagger}^{(i) } }
\end{displaymath}
in which $H_2$ has already been solved in (\ref{coefs}). Since both $H_0$ and $H_2$ are linear transformations, $H_1=H_0\cdot H_2^{-1}$ is linear as well.
In practice, we will firstly calculate
$H_1$ based on the unitarity of superconformal partial waves and then solve $H_0$ in terms of $H_1$ and $H_2$.

The transformation $H_1$ has been solved in Appendix C, and the most general $\cN=1$ superconformal block $\cG_{\De,\ell}^{r,\tl r}$
is written in terms of $\la_{\Phi_1\Phi_2\cO}^{(i)}$ and $(\la_{\Phi_4^\dagger\Phi_3^\dagger\cO}^{(i)})^*$.
Transformation from $(\la_{\Phi_4^\dagger\Phi_3^\dagger\cO}^{(i)})^*$ to $\la_{\Phi_3\Phi_4\cO^\dagger}^{(i)}$
has been solved in (\ref{coefs}), its inverse map gives $H_2(\tl r)$:
\bea
\left( \begin{array}{c}
(\la_{\Phi_4^\dagger\Phi_3^\dagger\cO}^{(0)})^* \\
(\la_{\Phi_4^\dagger\Phi_3^\dagger\cO}^{(2)})^* \\
(\la_{\Phi_4^\dagger\Phi_3^\dagger\cO}^{(1)})^* \\
(\la_{\Phi_4^\dagger\Phi_3^\dagger\cO}^{(3)})^*
\end{array} \right)
& = &
\left( \begin{array}{cccc}
1~ & ~0~ & ~0~ &~ 0 \\
\frac{1}{2}\tl r^2 & 1 & \tl r & \frac{1}{2}\tl r \\
\tl r & 0 & 1 & 0 \\
0 & 0 & 0 & 1
\end{array} \right)
\left( \begin{array}{c}
\la_{\Phi_3\Phi_4\cO^\dagger}^{(0)} \\
\la_{\Phi_3\Phi_4\cO^\dagger}^{(2)} \\
\la_{\Phi_3\Phi_4\cO^\dagger}^{(1)} \\
\la_{\Phi_3\Phi_4\cO^\dagger}^{(3)}
\end{array} \right) ,
\eea
and it satisfies
\beq
H_2(r)\cdot H_2(-r)=I_{4\times4},
\eeq
which is expected since the coefficients are invariant by taking complex conjugate twice.

It is straightforward to get transformation $H_0$ by combining the results of $H_1$ and $H_2$.
Here we do not present the explicit expression of $H_0$. The $\cN=1$ superconformal block is
\beq
\cG_{\De,\ell}^{r,\tl r}=a_1\, g_{\De,\ell}^{r,\tl{r}}+ a_2\, g_{\De+1,\ell+1}^{r,\tl{r}}+ a_3\, g_{\De+1,\ell-1}^{r,\tl{r}}
+ a_4\, g_{\De+2,\ell}^{r,\tl{r}},
\eeq
in which the coefficients of individual conformal blocks $a_i$ are written in terms of
$\la_{\Phi_1\Phi_2\cO}^{(i)}$ and $\la_{\Phi_3\Phi_4\cO^\dagger}^{(i)}$
{\small
\bea
a_1&=& \la_{\Phi_1 \Phi_2\cO}^{(0)} \la_{\Phi_3\Phi_4\cO^\dagger}^{(0)}, \label{aaa1} \\
a_2&=&\frac{\Delta +\ell }{(\Delta +\ell +1) (\Delta -R+\ell ) (\Delta +R+\ell )}
\left(\la_{\Phi_1\Phi_2\cO}^{(1)}+  \frac{r (\Delta -R+\ell )}{2 (\Delta +\ell )}\la_{\Phi_1\Phi_2\cO}^{(0)} \right)  \nonumber \\
&&\times\left(\la_{\Phi_3\Phi_4\cO^\dagger}^{(1)} +\frac{\tl r (\Delta +R+\ell )}{2 (\Delta +\ell)}
\la_{\Phi_3\Phi_4\cO^\dagger}^{(0)}\right), \label{aaa2} \\
a_3&=&\frac{\ell +2-\Delta }{(-\Delta +\ell +1) (-\Delta -R+\ell +2) (-\Delta +R+\ell +2)}  \nonumber \\
&&\times\left(\la_{\Phi_1\Phi_2\cO}^{(1)}+ \frac{\ell+1}{\ell}\la_{\Phi_1\Phi_2\cO}^{(3)}+\frac{r (-\Delta +R+\ell +2)}{2 (-\Delta +\ell +2)}\la_{\Phi_1\Phi_2\cO}^{(0)} \right) \nonumber\\
&&\times\left((\la_{\Phi_3\Phi_4\cO^\dagger}^{(1)} +\frac{\ell+1}{\ell}\la_{\Phi_3\Phi_4\cO^\dagger}^{(3)}+\frac{\tl r (-\Delta -R+\ell +2)}{2 (-\Delta +\ell +2)}\la_{\Phi_3\Phi_4\cO^\dagger}^{(0)}\right),  \label{aaa3}\\
a_4 &=&\frac{4 (\Delta -1)^2 (-\Delta +\ell +2) (\Delta +\ell )}{\Delta ^2 (\ell +1-\Delta) (\Delta +\ell +1) (\ell +2-R-\Delta) (\ell +2+R-\Delta) (\Delta -R+\ell ) (\Delta +R+\ell )} \times\nonumber\\
&&\left[ -\frac{(\Delta -R+\ell ) \left(R \left(\ell  (\ell +2)-\Delta  \left(\Delta +r^2-2\right)\right)+(\ell +2-\Delta) \left((\Delta +\ell )^2-\Delta  r^2\right)\right)}{8 (\Delta -1) (\ell +2-\Delta) (\Delta +\ell )}
\la_{\Phi_1\Phi_2\cO}^{(0)} \right. \nonumber \\
&&\left.
+ \la_{\Phi_1\Phi_2\cO}^{(2)}+\frac{r (\Delta  (R+2-\Delta)+\ell  (\ell +2))}{2 (\ell +2-\Delta ) (\Delta +\ell )}\la_{\Phi_1\Phi_2\cO}^{(1)}+\frac{r (\Delta -R+\ell )}{4 (\Delta +\ell )}\la_{\Phi_1\Phi_2\cO}^{(1)}\right]\times  \nonumber \\
&&\left[ \frac{(\Delta +R+\ell ) \left(R \left(\ell  (\ell +2)-\Delta  \left(\Delta +\tl r^2-2\right)\right)-(\ell +2-\Delta) \left((\Delta +\ell )^2-\Delta  \tl r^2\right)\right)}{8 (\Delta -1) (\ell +2-\Delta ) (\Delta +\ell )}
\la_{\Phi_3\Phi_4\cO^\dagger}^{(0)}  \right. \nonumber \\
&&\left.
+ \la_{\Phi_3\Phi_4\cO^\dagger}^{(2)}+\frac{\tl r (\Delta  (-R+2-\Delta)+\ell  (\ell +2))}{2 (\ell +2-\Delta) (\Delta +\ell )}\la_{\Phi_3\Phi_4\cO^\dagger}^{(1)}+\frac{\tl r (\Delta +R+\ell )}{4 (\Delta +\ell )}\la_{\Phi_3\Phi_4\cO^\dagger}^{(3)}\right]. \label{aaa4}
\eea }
Comparing with the superconformal blocks (\ref{aa1}-\ref{aa4}) in terms of $(\la_{\Phi_4^\dagger\Phi_3^\dagger\cO}^{(i)})^*$, above superconformal blocks
show improved symmetry that terms appear in pairs with correspondences
\beq
\la_{\Phi_1\Phi_2\cO}^{(i)} \leftrightarrow \la_{\Phi_3\Phi_4\cO^\dagger}^{(i)},~~r\leftrightarrow \tl r,~~R\leftrightarrow -R.
\eeq

Taking $r=\tl r=R=0$, the coefficients $a_i$ presented in (\ref{aaa1}-\ref{aaa4}) reduce to the results obtained in \cite{Khandker:2014mpa}.
For non-vanishing $r,~\tl r, R$, if certain fields $\Phi$'s in four-point function satisfy shortening conditions, like chirality,
the tensor structures can be simplified and
there will be strong constraints on the coefficients $\la_{\Phi_i\Phi_2\cO}^{(i)}$. In this case the superconformal blocks can be
conveniently solved through superconformal Casimir approach \cite{Fitzpatrick:2014oza, Bobev:2015jxa, Lemos:2015awa}.
As a non-trivial check, we compare our work with previous results on $\cN=1$ superconformal blocks obtained from
superconformal Casimir approach \cite{Bobev:2015jxa, Lemos:2015awa}.

In \cite{Bobev:2015jxa} superconformal blocks in SCFTs with four supercharges have been studied. The authors considered four-point function
$\<\Phi_1(1)X_1(2,\bar{2})\Phi_2(3)X_2(4,\bar{4})\>$, in which $\Phi_{1,2}$ are chiral, while $X_{1,2}$ are scalars with arbitrary superconformal weights.
As shown in (\ref{chiX}), chirality conditions of $\Phi_1$ and $\Phi_2$ lead to following constraints on the coefficients
\bea
(\la_{\Phi_1X_1\cO}^{(0)},~\la_{\Phi_1X_1\cO}^{(2)},~\la_{\Phi_1X_1\cO}^{(1)},~\la_{\Phi_1X_1\cO}^{(3)})&=&
\la_{\Phi_1X_1\cO} (1, ~e_1(2e_1-\ell),~-2e_1,~\ell),  \\
(\la_{\Phi_2X_2\cO^\dagger}^{(0)},~\la_{\Phi_2X_2\cO^\dagger}^{(2)},~\la_{\Phi_2X_2\cO^\dagger}^{(1)},~\la_{\Phi_2X_2\cO^\dagger}^{(3)})&=&
\la_{\Phi_2X_2\cO^\dagger} (1, ~e_2(2e_2-\ell),~-2e_2,~\ell),
\eea
where parameters $e_1$ and $e_2$ are
\beq
e_1=\frac{1}{4}(\De+\ell+2r+R),~~~~e_2=\frac{1}{4}(2-\De+\ell+2\tl r-R),
\eeq
and here the scaling dimension differences $r$ and $\tl r$ become $r=\De_{\Phi_1}-\De_{X_1},~\tl r=\De_{\Phi_2}-\De_{X_2}$. Plugging these constraints in (\ref{aaa1}-\ref{aaa4}), coefficients of conformal blocks in $\cG_{\De,\ell}^{r,\tl r}$ turn into
\bea
a_1&=& \la_{\Phi_1X_1\cO} \la_{\Phi_2X_2\cO^\dagger},\\
a_2&=& \frac{ (\Delta +r+\ell ) (\Delta +\tl r+\ell )}{4 (\Delta +\ell ) (\Delta +\ell +1)}\la_{\Phi_1X_1\cO} \la_{\Phi_2X_2\cO^\dagger}, \\
a_3&=& \frac{(\Delta +r-\ell -2) (\Delta +\tl r-\ell -2)}{4 (-\Delta +\ell +1) (-\Delta +\ell +2)}\la_{\Phi_1X_1\cO} \la_{\Phi_2X_2\cO^\dagger},\\
a_4&=& \frac{(\Delta +r-\ell -2) (\Delta +\tl r-\ell -2)  (\Delta +r+\ell )(\Delta +\tl r+\ell )}{16 (-\Delta +\ell +1) (-\Delta +\ell +2) (\Delta +\ell ) (\Delta +\ell +1)} \la_{\Phi_1X_1\cO} \la_{\Phi_2X_2\cO^\dagger},
\eea
which are in agreement with the results obtained in \cite{Bobev:2015jxa}. $\cN=1,~2$ superconformal blocks are also presented in
\cite{Lemos:2015awa}, in which the four-point function consists of chiral-antichiral scalars with arbitrary $U(1)$ R-charges.
For $\cN=1$ case, the superconformal blocks are similar to above expressions and are well consistent with our results.

\section{Discussion}
In this work we have computed the most general $\cN=1$ superconformal partial waves $\cW_\cO\propto\<\Phi_1\Phi_2|\cO|\Phi_3\Phi_4\>$, in which the scalars $\Phi_i$ have arbitrary scaling dimensions and $U(1)$ R-charges. Our computations are based on the superembedding space formalism and supershadow approach, which provide a systematic way to study $\cN=1$ superconformal blocks.
Unitarity of SCFTs has been used to evaluate the coefficients in the three-point function of supershadow operator. Besides, it shows deep connections between conformal field theories and mathematical properties of hypergeometric functions throughout the computations.
Our results nicely reproduce all the known results on the $\cN=1$ superconformal blocks under certain restrictions.

The superconformal blocks of operators with arbitrary scaling dimensions and R-charges are crucial ingredients for the mixed operator conformal bootstrap, and our results provide necessary materials for bootstrapping any $\cN=1$ SCFTs. An attractive problem
is the $4D$ $\cN=1$ minimal SCFT, which has no Lagrangian description  and its existence is only revealed in superconformal bootstrap \cite{Poland:2011ey, Poland:2015mta}. More details of the theory are expected to be studied through bootstrapping the mixed operator correlators \cite{Li:toa}. Our current results on the SCFTs are limited to $4D$ $\cN=1$ scalars, and obviously it can be generalized from three aspects: dimension of spacetime, number of supercharges and spin of the fields in four-point correlator. The supershadow approach has impressive successes in solving $4D$ $\cN=1$ scalar superconformal blocks, we hope this method, and its generalizations can
be used to obtain the superconformal blocks of spinning operators in other dimensional spacetime with different supercharges.

\section*{Acknowledgements}
We are grateful to Daniel Robbins and Junchen Rong for useful discussions. We would like to thank David Simmons-Duffin for valuable comments on this project. Z. L especially wants to thank Madalena Lemos and Pedro Liendo for their enlightening discussions and sharing their {\it Mathematica} code on evaluating hypergeometric functions, which is extremely helpful for our computations.


\newpage
\appendix
\section{Gegenbauer Polynomial and Some Identities}
It has been shown in \cite{Dolan:2000ut, SimmonsDuffin:2012uy, Fitzpatrick:2014oza, Khandker:2014mpa} that $\cN_\ell$ appearing in the superconformal/conformal partial wave integration directly relates to Gegenbauer polynomial $C_{\ell}^{(\la)}(x)$
\beq
\cN_{\ell}\equiv\left(\bar{\cS}1\bar{2}\cS\right)^{\ell}\overleftrightarrow{\mathcal{D}_{\ell}}\left(\bar{\cT}3\bar{4}\cT\right)^{\ell}
=\frac{1}{\ell!^{2}}\left(\partial_{\cS}0\partial_{\cT}\right)^{\ell}\left(\cS\bar{2}1\bar{0}3\bar{4}\cT\right)^{\ell}
=\left(-1\right)^{\ell}y^{\frac{\ell}{2}}C_{\ell}^{(1)}(x) ,
\eeq
in which
\beq
x\equiv\frac{\langle \bar{2}1\bar{0}3\bar{4}0\rangle }{2\sqrt{y}},~~~~ y\equiv\frac{1}{2^{6}}\langle \bar{0}1\rangle \langle \bar{2}0\rangle \langle \bar{0}3\rangle \langle \bar{4}0\rangle \langle \bar{2}1\rangle \langle \bar{4}3\rangle. \label{xy}
\eeq
Giving $\theta_{\rm ext}=0$, variables $x$ and $y$  turn into
\bea
x&\longrightarrow& x_{0}\equiv-\frac{X_{13}X_{20}X_{40}}{2\sqrt{X_{10}X_{20}X_{30}X_{40}X_{12}X_{34}}}-\left(1\leftrightarrow2\right)-\left(3\leftrightarrow4\right), \label{t0} \\
y&\longrightarrow& y_{0}\equiv\frac{1}{2^{12}}X_{10}X_{20}X_{30}X_{40}X_{12}X_{34},\label{s0}
\eea
in which the supertraces $\<i\bar j\>$ have been reduced to inner products of six dimensional vectors $X_{ij}$. Besides we follow the conventions used in  \cite{Khandker:2014mpa} that the super-parameters are replaced by
\beq
\cS\rightarrow S,~\bar\cS\rightarrow \bar S,~ \cN_\ell\rightarrow N_\ell
\eeq
after setting $\theta_{\rm ext}=0$, and the Gegenbauer polynomial  $N_\ell$ reads
\beq
N_{\ell}=(\bar{S}1\bar{2}S)^{\ell}\overleftrightarrow{\mathcal{D}_{\ell}}(\bar{T}3\bar{4}T)^{\ell}=\frac{1}{\ell!^{2}}(\partial_{S}0\partial_{T})^{\ell}(S\bar{2}1\bar{0}3\bar{4}T)^{\ell} .\label{Nl}
\eeq

 Giving $0=\bar0$, one can show
 \begin{equation}
S\bar{2}1\bar{0}3\bar{4}T=\frac{1}{4}X_{10}S\bar{2}3\bar{4}T-\frac{1}{4}X_{20}S\bar{1}3\bar{4}T
=\frac{1}{4}X_{30}S\bar{2}1\bar{4}T-\frac{1}{4}X_{40}S\bar{2}1\bar{3}T \label{expand}
\end{equation}
based on the Clifford algebra and the transverse conditions of auxiliary fields $S\bar0=\bar0T=0$. It clearly shows that $S\bar{2}1\bar{0}3\bar{4}T$
is antisymmetric under $1\leftrightarrow2$ or $3\leftrightarrow4$.

Let us consider following formulas related to the Gegenbauer polynomials
\bea
\left(\partial_{S}0\partial_{T}\right)^{\ell}\left(S\bar{2}1\bar{0}3\bar{4}T\right)^{\ell-1}\left( X_{10}S\bar{2}3\bar{4}T + X_{20}S\bar{1}3\bar{4}T + X_{10}S\bar{2}4\bar{3}T + X_{20}S\bar{1}4\bar{3}T\right), \label{An}\\
\left(\partial_{S}0\partial_{T}\right)^{\ell}\left(S\bar{2}1\bar{0}3\bar{4}T\right)^{\ell-1}\left( X_{10}S\bar{2}3\bar{4}T + X_{20}S\bar{1}3\bar{4}T - X_{10}S\bar{2}4\bar{3}T - X_{20}S\bar{1}4\bar{3}T\right),  \label{Bn}\\
\left(\partial_{S}0\partial_{T}\right)^{\ell}\left(S\bar{2}1\bar{0}3\bar{4}T\right)^{\ell-1}\left( X_{10}S\bar{2}3\bar{4}T - X_{20}S\bar{1}3\bar{4}T + X_{10}S\bar{2}4\bar{3}T - X_{20}S\bar{1}4\bar{3}T\right), \label{Cn} \\
\left(\partial_{S}0\partial_{T}\right)^{\ell}\left(S\bar{2}1\bar{0}3\bar{4}T\right)^{\ell-1}\left( X_{10}S\bar{2}3\bar{4}T - X_{20}S\bar{1}3\bar{4}T - X_{10}S\bar{2}4\bar{3}T + X_{20}S\bar{1}4\bar{3}T\right), \label{Dn}
\eea
which are symmetric or anti-symmetric under coordinate interchanges $1\leftrightarrow2$ and $3\leftrightarrow4$.
These polynomials appear in the conformal integral (\ref{Int1}) from differentiations $(\partial_{\bar0}z)\cdot (\partial_{\bar0}N_\ell)$ or
$(\partial_{\bar0}\frac{1}{D_\ell})\cdot (\partial_{\bar0}N_\ell)$ and inherit the symmetry properties from tensor structure terms in (\ref{NF2}).
We need to find their close relationships with Gegenbauer polynomials to accomplish the conformal integration (\ref{Int1}).

Formulas in (\ref{An}) and (\ref{Dn}) are invariant under simultaneous coordinate interchange $1\leftrightarrow2,~3\leftrightarrow4$, and they can be easily
simplified into compact form $N_\ell$. Specifically, the formula (\ref{Dn}) gives
\beq
8\left(\partial_{S}0\partial_{T}\right)^{\ell}\left(S\bar{2}1\bar{0}3\bar{4}T\right)^{\ell}\propto N_\ell,
\eeq
while for (\ref{An}), one can show that it reduces to
\bea
&&\frac{1}{4}\left(\partial_{S}0\partial_{T}\right)^{\ell}\left(S\bar{2}1\bar{0}3\bar{4}T\right)^{\ell-1}\left( X_{10}X_{34}S\bar{2}T + X_{20}X_{34}S\bar{1}T \right) \nonumber \\
&&\hspace{20mm}=\frac{\ell(\ell+1)}{8}X_{10}X_{20}X_{34}\left(\partial_{S}0\partial_{T}\right)^{\ell-1}\left(S\bar{2}1\bar{0}3\bar{4}T\right)^{\ell-1}\nonumber\\
&&\hspace{20mm}\propto  X_{10}X_{20}X_{34} N_{\ell-1}.
\eea

In contrast, formulas in (\ref{Bn}) and (\ref{Cn}) are antisymmetric under $1\leftrightarrow2,~3\leftrightarrow4$. It is easy to show that formula (\ref{Cn}) vanishes.

Similarly, we can reduce following formulas to compact forms proportional to $N_\ell$:
\bea
\left(\partial_{S}0\partial_{T}\right)^{\ell}\left(S\bar{2}1\bar{0}3\bar{4}T\right)^{\ell-1}\left( X_{30}S\bar{2}1\bar{4}T + X_{40}S\bar{2}1\bar{3}T + X_{30}S\bar{1}2\bar{4}T + X_{40}S\bar{1}2\bar{3}T\right), \label{Am}\\
\left(\partial_{S}0\partial_{T}\right)^{\ell}\left(S\bar{2}1\bar{0}3\bar{4}T\right)^{\ell-1}\left( X_{30}S\bar{2}1\bar{4}T + X_{40}S\bar{2}1\bar{3}T - X_{30}S\bar{1}2\bar{4}T - X_{40}S\bar{1}2\bar{3}T\right),  \label{Bm}\\
\left(\partial_{S}0\partial_{T}\right)^{\ell}\left(S\bar{2}1\bar{0}3\bar{4}T\right)^{\ell-1}\left( X_{30}S\bar{2}1\bar{4}T - X_{40}S\bar{2}1\bar{3}T + X_{30}S\bar{1}2\bar{4}T - X_{40}S\bar{1}2\bar{3}T\right), \label{Cm} \\
\left(\partial_{S}0\partial_{T}\right)^{\ell}\left(S\bar{2}1\bar{0}3\bar{4}T\right)^{\ell-1}\left( X_{30}S\bar{2}1\bar{4}T - X_{40}S\bar{2}1\bar{3}T - X_{30}S\bar{1}2\bar{4}T + X_{40}S\bar{1}2\bar{3}T\right), \label{Dm}
\eea
except (\ref{Bm}).

The formulas (\ref{Bn}) and (\ref{Bm}) can not be simply written in terms of $N_\ell$, nevertheless, their relationships with the
Gegenbauer polynomials are given in the recursion equations,
which can be used to obtain the final results of conformal integrations they involve in.

Denote
\bea
R_\ell\equiv \frac{\ell}{{\ell!}^2} \left(\partial_{S}0\partial_{T}\right)^{\ell}\left(S\bar{2}1\bar{0}3\bar{4}T\right)^{\ell-1}\times&&\nonumber \\
 && \hspace{-40mm}  \left( X_{10}S\bar{2}3\bar{4}T + X_{20}S\bar{1}3\bar{4}T - X_{10}S\bar{2}4\bar{3}T - X_{20}S\bar{1}4\bar{3}T\right),  \label{Rl}\\
P_\ell\equiv \frac{\ell}{{\ell!}^2}  \left(\partial_{S}0\partial_{T}\right)^{\ell} \left(S\bar{2}1\bar{0}3\bar{4}T\right)^{\ell-1} \times&& \nonumber\\
&& \hspace{-40mm} \left( X_{30}S\bar{2}1\bar{4}T + X_{40}S\bar{2}1\bar{3}T - X_{30}S\bar{1}2\bar{4}T - X_{40}S\bar{1}2\bar{3}T\right),  \label{Pl}
\eea
and
\bea
\De_A \equiv  \frac{1}{8}(X_{20}X_{40}X_{13}-X_{20}X_{30}X_{14}+X_{10}X_{40}X_{23}-X_{10}X_{30}X_{24}), \\
\De_B \equiv \frac{1}{8}(X_{20}X_{40}X_{13}+X_{20}X_{30}X_{14}-X_{10}X_{40}X_{23}-X_{10}X_{30}X_{24}).
\eea
Note the sign differences among $x_0$, $\De_A$ and $\De_B$.
The crucial properties of $R_\ell$ and $P_\ell$ are that they satisfy the following mutual recursion relations:
\bea
R_\ell &=& \ell\De_A N_{\ell-1}+\frac{1}{2^6}X_{10}X_{20}X_{34}P_{\ell-1}, \\
P_\ell &=& \ell\De_B N_{\ell-1}+\frac{1}{2^6}X_{30}X_{40}X_{12}R_{\ell-1},
\eea
which leads to the independent recursion relations of $R_\ell$ and $P_\ell$:
\bea
R_\ell &=& \ell\De_A N_{\ell-1}+\frac{1}{2^6}(\ell-1)X_{10}X_{20}X_{34}\De_B N_{\ell-2}+y_0R_{\ell-2}, \label{Rlr}\\
P_\ell &=& \ell\De_B N_{\ell-1}+\frac{1}{2^6}(\ell-1)X_{30}X_{40}X_{12}\De_A N_{\ell-2}+y_0P_{\ell-2}. \label{Plr}
\eea
Above two recursion equations are needed to determine the conformal integrations of the antisymmetric terms
in (\ref{anti}).

\section{Superconformal Integrations of Symmetric Terms}
The superconformal partial waves $\cW_\cO$ are largely determined by the tensor structures in (\ref{NF2}). These terms
are separated into two parts: invariant and antisymmetric terms according to their transformations under coordinate interchange $1\leftrightarrow2,~3\leftrightarrow4$. Here we show the main steps toward
contributions of invariant terms on $\cW_\cO$. Due to the gauge adopted in (\ref{3p}), we can obtain the results
straightforwardly, similar to the steps used in  \cite{Khandker:2014mpa} but generalized to $\Phi_i$'s with arbitrary superconformal weights.

As discussed before, there are two steps to accomplish the superconformal integrations for $\cW_\cO$: partial derivatives and conformal integration.
The partial derivatives can be obtained by the same steps provided in \cite{Khandker:2014mpa} with coefficients replacements
\beq
\frac{\ell+\De}{2}\rightarrow 2\de,~~~~~~~\frac{2+\ell-\De}{2}\rightarrow 2\de'.
\eeq
The conformal integrations are modified accordingly, specifically there are new terms proportional to the scaling dimension differences $r,\tl r$:
\begin{eqnarray}
\int D^{4}X_{0}\left.\frac{X_{12}}{X_{10}X_{20}}\frac{N_{\ell}}{D_{\ell}}\right|_{\bar{0}=0}&=&\frac{c_{\ell}\;\xi_{\De+2,2-\De,\tl{r},\ell}}
{X_{12}^{\frac12(\De-\ell)}X_{34}^{-\frac12(\De+\ell-2)}}
\left(\frac{X_{24}}{X_{14}}\right)^{\frac{r}{2}}\left(\frac{X_{14}}{X_{13}}\right)^{\frac{\tl{r}}{2}}g_{\De+2,\ell}^{r,\tl{r}}(u,v) , \\
\int D^{4}X_{0}\left.\frac{X_{34}}{X_{30}X_{40}}\frac{N_{\ell}}{D_{\ell}}\right|_{\bar{0}=0}&=&\frac{c_{\ell}\;\xi_{\De,4-\De,\tl{r},\ell}}
{X_{12}^{\frac12(\De-\ell)}X_{34}^{-\frac12(\De+\ell-2)}}
\left(\frac{X_{24}}{X_{14}}\right)^{\frac{r}{2}}\left(\frac{X_{14}}{X_{13}}\right)^{\frac{\tl{r}}{2}}g_{\De,\ell}^{r,\tl{r}}(u,v), \\
\int D^{4}X_{0}\left. X_{12}X_{34}\frac{N_{\ell-1}}{D_{\ell}}\right|_{\bar{0}=0}&=&
\frac{c_{\ell-1}\;\xi_{\De+1,3-\De,\tl{r},\ell-1}}
{X_{12}^{\frac12(\De-\ell)}X_{34}^{-\frac12(\De+\ell-2)}}
\left(\frac{X_{24}}{X_{14}}\right)^{\frac{r}{2}}\left(\frac{X_{14}}{X_{13}}\right)^{\frac{\tl{r}}{2}}g_{\De+1,\ell-1}^{r,\tl{r}}(u,v)  ,
\end{eqnarray}
\vspace{-\baselineskip}
\begin{eqnarray}
&&\hspace{-5mm} \int D^{4}X_{0}\left.\left[\frac{X_{13}}{X_{10}X_{30}}+\frac{X_{23}}{X_{20}X_{30}}+\frac{X_{14}}{X_{10}X_{40}}+\frac{X_{24}}{X_{20}X_{40}}
\right]\frac{N_{\ell}}{D_{\ell}}\right|_{\bar{0}=0}=\nonumber\\
&&\hspace{10mm}
\frac{c_{\ell}\,\xi_{\De+1,3-\De,1+\tl{r},\ell}}{X_{12}^{\frac12(\De-\ell)}X_{34}^{-\frac12(\De+\ell-2)}}
\left(\frac{X_{24}}{X_{14}}\right)^{\frac{r}{2}}\left(\frac{X_{14}}{X_{13}}\right)^{\frac{\tl{r}}{2}}
\left[
\frac{4 \left(\tilde{r}^2+(\Delta -\ell -2) (\Delta +\ell )\right)}{\left(\tilde{r}+\Delta -\ell -2\right) \left(\tilde{r}+\Delta +\ell \right)}
 g_{\De,\ell}^{r,\tl{r}} \right.\nonumber\\
&&\hspace{15mm}
\left.+\frac{\left(r^2+(\Delta -\ell -2) (\Delta +\ell )\right) \left(\tilde{r}-\Delta -\ell \right) \left(\tilde{r}-\Delta +\ell +2\right)}{4 (\Delta -\ell -2) (\Delta -\ell -1) (\Delta +\ell ) (\Delta +\ell +1)}
g_{\De+2,\ell}^{r,\tl{r}} \right. \nonumber\\
&&\hspace{15mm}
\left.+\frac{r \tilde{r} \left(\tilde{r}-\Delta -\ell \right)}{(\Delta +\ell ) (\Delta +\ell +1) \left(\tilde{r}+\Delta -\ell -2\right)}
g_{\De+1,\ell+1}^{r,\tl{r}} \right.  \nonumber\\
&&\hspace{15mm}
\left.+\frac{r \tilde{r} \left(\tilde{r}-\Delta +\ell +2\right)}{(\Delta -\ell -2) (\Delta -\ell -1) \left(\tilde{r}+\Delta +\ell \right)}
g_{\De+1,\ell-1}^{r,\tl{r}}\right], \\
&&\hspace{-5mm}\int D^{4}X_{0}\left.\left[\frac{X_{13}}{X_{10}X_{30}}-\frac{X_{23}}{X_{20}X_{30}}-\frac{X_{14}}{X_{10}X_{40}}+\frac{X_{24}}{X_{20}X_{40}}
\right]\frac{N_{\ell}}{D_{\ell}}\right|_{\bar{0}=0} = \nonumber\\
&&\hspace{10mm}
\frac{c_{\ell}\,\xi_{\De+1,3-\De,1+\tl{r},\ell}}{X_{12}^{\frac12(\De-\ell)}X_{34}^{-\frac12(\De+\ell-2)}}
\left(\frac{X_{24}}{X_{14}}\right)^{\frac{r}{2}}\left(\frac{X_{14}}{X_{13}}\right)^{\frac{\tl{r}}{2}}
\left[
\frac{(\Delta -\ell -2) \left(-\tilde{r}+\Delta +\ell \right)}{(\Delta +\ell +1) \left(\tilde{r}+\Delta -\ell -2\right)} g_{\De+1,\ell+1}^{r,\tl{r}}\right.  \nonumber\\
&&\hspace{15mm}
\left.+\frac{(\Delta +\ell ) \left(-\tilde{r}+\Delta -\ell -2\right)}{(\Delta -\ell -1) \left(\tilde{r}+\Delta +\ell \right)} g_{\De+1,\ell-1}^{r,\tl{r}}\right].
\end{eqnarray}

\section{Solution of the Shadow Coefficients Transformation}
Here we solve the linear transformation $H_1$ between the supershadow coefficients $\la_{\Phi_2^\dagger\Phi_1^\dagger\tl\cO}^{(i)}$ and
$(\la_{\Phi_1 \Phi_2\cO}^{(i)})^*$. As proposed in
\cite{Khandker:2014mpa}, the unitarity of superconformal partial wave plays a crucial role in determining $H_1$.

The linear transformation $H_1$ is described by a $4\times4$ matrix
\bea
\left( \begin{array}{c}
\la_{\Phi_2^\dagger\Phi_1^\dagger\tl\cO}^{(0)} \\
\la_{\Phi_2^\dagger\Phi_1^\dagger\tl\cO}^{(2)} \\
\la_{\Phi_2^\dagger\Phi_1^\dagger\tl\cO}^{(1)} \\
\la_{\Phi_2^\dagger\Phi_1^\dagger\tl\cO}^{(3)}
\end{array} \right)
& = &
\left( \begin{array}{cccc}
a ~&~ b~ &~ e~ &~ g \\
c & d & f & h \\
u & v & p & k \\
w & t & q & s
\end{array} \right)
\left( \begin{array}{c}
(\la_{\Phi_1 \Phi_2\cO}^{(0)})^* \\
(\la_{\Phi_1 \Phi_2\cO}^{(2)})^* \\
(\la_{\Phi_1 \Phi_2\cO}^{(1)})^* \\
(\la_{\Phi_1 \Phi_2\cO}^{(3)})^*
\end{array} \right) .
\eea
Note that in \cite{Khandker:2014mpa}  the $4\times4$ matrix is block diagonal protected by the parity of coefficients under coordinate exchange in the three-point function.
While for the three-point function with general superconformal weights, the coordinate exchange symmetry is broken by arbitrary superconformal weights,
therefore in our case the $4\times4$ matrix is not simply block diagonal, nevertheless the unitarity, together with extra constraint is still useful to solve the transformation $H_1$.

Giving $\Phi_3=\Phi_2^\dagger$ and $\Phi_4=\Phi_1^\dagger$, unitarity requires that the four coefficients $a_i$ of conformal blocks appearing in the superconformal blocks $\cG_{\De,\ell}^{r,\tl r}$ are positive. By transforming coefficients $\la_{\Phi_2^\dagger\Phi_1^\dagger\tl\cO}^{(i)}$ to $(\la_{\Phi_1 \Phi_2\cO}^{(i)})^*$,
this is equivalent to the following equations:
\bea
\left(-\de' \left[\frac{(2-\Delta ) \tl r^2 -(\ell +2-\Delta) (\Delta +\ell )}{(\Delta -1) ( \ell +2-\Delta) (\Delta +\ell )}
\right.\right.&&\left.\left.\hspace{-9mm}
2\delta +1\right],
~1,
~\frac{\tl r ((\Delta -2) R+(-\Delta +\ell +2) (\Delta +\ell ))}{2 (-\Delta +\ell +2) (\Delta +\ell )},\right. \nonumber\\
\left.
\frac{\tl r (R+\ell +2-\Delta)}{4 (-\Delta +\ell +2)}\right)\cdot H_1  &\propto &(1,~0,~0,~0),  \label{ueq1}\\
\left(\frac{\tl r (R+\ell +2-\Delta )}{2 (-\Delta +\ell +2)}, ~0, ~1, ~0\right)\cdot H_1 &\propto &\left(\frac{r (\Delta -R+\ell )}{2 (\Delta +\ell )}, ~0, ~1, ~0\right),  \label{ueq2}\\
\left(\frac{\tl r (\Delta -R+\ell )}{2 (\Delta +\ell )}, ~0,~1,~\frac{\ell+1}{\ell}\right)\cdot H_1 &\propto &
\left(\frac{r (-\Delta +R+\ell +2)}{2 (-\Delta +\ell +2)}, ~0,~1,~\frac{\ell+1}{\ell}\right), \label{ueq3} \\
(1,~0,~0,~0)\cdot H_1 &\propto&\left( -\delta  \left(1- 2 \de'
\frac{ r^2 \Delta  -(\Delta +\ell ) (-\Delta +\ell +2)}{(\Delta -1) (\ell +2-\Delta) (\Delta +\ell )}\right), \right. \nonumber\\
&& \hspace{-28mm} \left.~1,~ \frac{r (\Delta  (-\Delta +R+2)+\ell  (\ell +2))}{2 (-\Delta +\ell +2) (\Delta +\ell )},
~\frac{r (\Delta -R+\ell )}{4 (\Delta +\ell )} \right),  \label{ueq4}
\eea
in which $\tl r=-r$. From above equation groups we can solve $15$ out of $16$ $H_1$'s elements (except $c$) up to three re-scaling coefficients.

Then we consider two three-point functions $\<\Phi X \cO\>$ and $\<X\Phi^\dagger \tl\cO\>$, in which $\Phi:(0,~0,~q_1,~0)$ is a chiral field while
$X: (0,~0,~q_2,~q_2)$ is real \footnote{X could be any scalar and the results will be the same, here we set X as real for convenience.}. Such kind of three-point function has been studied
in (\ref{chiX}). Due to the chirality of $\Phi$, the four coefficients actually satisfy the constraint
\bea
(\la_{\Phi X\cO}^{(0)},~ \la_{\Phi X\cO}^{(2)}, ~\la_{\Phi X\cO}^{(1)},~ \la_{\Phi X\cO}^{(3)})&=&\la_{\Phi X\cO}(1,~ \de(2\de-\ell),~-2\de,~ \ell),
\label{3peff1} \\
(\la_{X\Phi^\dagger\tl\cO}^{(0)},~ \la_{X\Phi^\dagger\tl\cO}^{(2)}, ~\la_{X\Phi^\dagger\tl\cO}^{(1)},
~ \la_{X\Phi^\dagger\tl\cO}^{(3)})&=&\la_{X\Phi^\dagger\tl\cO}(1,~ \de'(2\de'-\ell),~-2\de',~ \ell), \label{3peff2}
\eea
in which $\de=\frac{\De+\ell+R+2r}{4} $ and $\de'=\frac{2-\De+\ell+R}{4}$ with $R=-q_1$, $r=q_1-2q_2$. then the transformation between coefficients in (\ref{3peff2}) and the complex conjugate of (\ref{3peff1}) gives
\bea
\left( \begin{array}{c}
1 \\
\de'(2\de'-\ell) \\
-2\de' \\
\ell
\end{array} \right)
& \propto &
\left( \begin{array}{cccc}
a ~&~ b~ &~ e~ &~ g \\
c & d & f & h \\
u & v & p & k \\
w & t & q & s
\end{array} \right)
\left( \begin{array}{c}
1 \\
\de(2\de-\ell) \\
-2\de \\
\ell
\end{array} \right) . \label{ueq5}
\eea
Plugging the solutions of equation groups (\ref{ueq1}-\ref{ueq4}) into (\ref{ueq5}), we can solve all the $16$ elements in $H_1$ and three
re-scaling coefficient up to the re-scaling coefficient of (\ref{ueq5}), denoted as $z_*$. The results are
\bea
{\small
\alpha_* \times \left(
\begin{array}{cccc}
 a_* & -\frac{8 (\ell -\Delta +2) (\ell +\Delta )}{\ell +\Delta-R } & -\frac{4 r (\ell  (\ell +2)+(R-\Delta +2) \Delta )}{\ell +\Delta-R } & -2 r (\ell -\Delta +2) \\
 c_* & d_* & f_* & h_* \\
 u_* & -\frac{4 r (R+\ell -\Delta +2) (\ell +\Delta )}{ \ell +\Delta-R } & p_* & -r^2 (R+\ell -\Delta +2) \\
 w_* & \frac{8 r R \ell }{ \ell +\Delta -R} & \frac{4 \ell  \left(R r^2+(\ell -\Delta +2) \Delta  (\ell +\Delta )\right)}{ \ell +\Delta-R } & s_* \\
\end{array}
\right)
}
\eea
in which the elements with long expressions are abbreviated as
{\scriptsize
\bea
\alpha_*&=&\frac{z_*(\Delta -1) (\Delta -R+\ell )}{\Delta  (-\Delta -r+\ell +2) (\Delta +r+\ell ) (-\Delta -R+\ell +2) (\Delta +R+\ell )}, \\
a_*&=&\frac{R \left(\ell  (\ell +2)-\Delta  \left(\Delta +r^2-2\right)\right)+(-\Delta +\ell +2) \left((\Delta +\ell )^2-\Delta  r^2\right)}{\Delta -1}, \\
d_*&=&\frac{R+\ell +2-\Delta }{(\Delta -1) (R-\ell-\Delta  )}\left(\Delta ^2 \left(r^2-R+\ell +4\right)-\Delta ^3 +\Delta  \left(\left(2-r^2\right) R+\ell  \left(r^2+\ell \right)-4\right)+\ell  (\ell +2) (R-\ell -2)\right), \\
h_*&=&\frac{r (R+\ell +2-\Delta )}{4 (\Delta -1)}\left(-\Delta  \left(\Delta ^2+\Delta -2 r^2-4\right)-(\Delta -1) R (\Delta +\ell +2)-(\Delta +1) \ell ^2-2 ((\Delta -1) \Delta +2) \ell -4\right), \\
u_*&=& -\frac{r (R+\ell +2-\Delta )}{2 (-1 + \De)}(-\Delta  \left((\Delta -3) \Delta -2 r^2+4\right)+(\Delta -1) R (\ell -\Delta )-(\Delta +1) \ell ^2+2 (\Delta -3) \Delta  \ell),\\
p_*&=&\frac{ R+\ell +2-\Delta}{(\Delta -1) (R-\ell -\Delta )}\left(r^2 (\Delta  (3 \Delta -R-2)+(3 \Delta -2) \ell )+\Delta  (-\Delta +\ell +2) (\Delta +\ell ) (\Delta +R-\ell -2)\right),\\
s_*&=&\frac{1}{\De-1}\left(r^2 ((\Delta -2) R+\Delta  (\Delta -\ell -2))+\Delta  (-\Delta +\ell +2) (\Delta +\ell ) (\Delta +R+\ell )\right), \\
w_*&=&r \ell  (4 \Delta +(R+\ell -\Delta ) (R+2 \Delta )),
\eea
}
and
{\scriptsize
\bea
c_*&=& \frac{\Delta +R-\ell -2}{8 (\Delta -1) (\ell +2-\Delta -R) (\Delta -R+\ell )} \left( 4 (\Delta -1) \Delta  r^2 R^3+(\Delta -1) R^4 (\ell +2-\Delta) (\Delta +\ell )  -4 (\Delta -1) \Delta  r^2  R  \right. \nonumber\\
 && \left.
\left((\Delta -4) \Delta -2 r^2+3 \ell  (\ell +2)+6\right)+2 R^2 (\ell  (\ell +2) (\ell  (\ell +2)+2)+
\Delta ^5-5 \Delta ^4-2 \Delta ^3 \left(r^2-5\right)
+2 \Delta ^2 \left(r^2-5\right)\right. \nonumber\\
&&\left.
-\Delta  \left(2 r^4-2 r^2 (\ell +1)^2+\ell  (\ell +2) (\ell  (\ell +2)+2)-4\right)+(\ell +2-\Delta) (\Delta +\ell ) \left(\Delta ^5-5 \Delta ^4
-2 \Delta ^3
\left(2 r^2+\ell  (\ell +2)  \right.\right.\right. \nonumber \\
&& \left.\left.\left.
-4\right)+2 \Delta ^2 \left(6 r^2+3 \ell  (\ell +2)-2\right)+\Delta  \left(4 r^4-4 r^2 (\ell  (\ell +2)+3)+\ell ^3 (\ell +4)-8 \ell \right)-\ell ^2 (\ell +2)^2\right)\right), \\
f_*&=& \frac{R+\ell +2-\Delta }{2 (\Delta -1) (\Delta -R+\ell )}\left(\ell  (\ell +2) (-R+\ell +2)+\Delta  \left(R \left(2 r^2-\ell  (\ell +3)-4\right)+\ell  \left(-2 r^2+\ell ^2+\ell +4\right)+R^2+4\right) \right.\nonumber \\
&&\left.+\Delta ^2 \left(-2 r^2-R^2+R (\ell +2)+(\ell -3) \ell \right)+\Delta ^3 (\ell -3)+\Delta ^4 \right).
\eea
}
The transformation $H_1$ presented above seems to be rather cumbersome, however it does satisfy following simple relation
\beq
H_1(\De,~R,~r)\cdot H_1(\De\rightarrow2-\De,~R\rightarrow-R,~r\rightarrow-r) \propto I_{4\times4}, \label{peq}
\eeq
which is expected since by applying the supershadow transformation twice we go back to the original coefficients.
Setting the Eq.~(\ref{peq}) to be strictly equal,  the overall coefficient $z_*$ can be fixed up to a factor $z_x$ satisfying
\beq
z_x(\De,~R,~r)\cdot z_x(\De\rightarrow2-\De,~R\rightarrow-R,~r\rightarrow-r)=1,
\eeq
which, however, has no effect on the superconformal block functions.

Besides the three-point correlators $\<\Phi X \cO\>$ and $\<X\Phi^\dagger\cO\>$, we can also partially fix the coefficients in the three-point correlators like
$\<\Phi^\dagger X \cO\>$, $\<X\Phi \cO\>$ and their supershadow duals. Their coefficients are expected to be related to the shadow coefficients by $H_1$ with proper redefinitions of parameter $r$ and $R$.  One can show that indeed above solution of $H_1$ can realize the transformation of shadow coefficients
with parameters $R\rightarrow-R$ and $r\rightarrow -r$, respectively.

Under
 transformation $H_1$, the coefficients $\la_{\Phi_3 \Phi_4\tl\cO}^{(i)}$ in (\ref{a1}-\ref{a4}) can be mapped to
$(\la_{\Phi_4^\dagger\Phi_3^\dagger\cO}^{(i)})^*$,
and now we are ready to write down the most general $\cN=1$ superconformal block $\cG_{\De,\ell}^{r,\tl r}$ in terms of three-point coefficients
$\la_{\Phi_1 \Phi_2\cO}^{(i)}$ and $(\la_{\Phi_4^\dagger\Phi_3^\dagger\cO}^{(i)})^*$ :
\beq
\cG_{\De,\ell}^{r,\tl r}=a_1\, g_{\De,\ell}^{r,\tl{r}}+ a_2\, g_{\De+1,\ell+1}^{r,\tl{r}}+ a_3\, g_{\De+1,\ell-1}^{r,\tl{r}}
+ a_4\, g_{\De+2,\ell}^{r,\tl{r}},
\eeq
where the coefficients of individual conformal blocks $a_i$ are
{\small
\bea
a_1&=& \la_{\Phi_1 \Phi_2\cO}^{(0)} (\la_{\Phi_4^\dagger\Phi_3^\dagger\cO}^{(0)})^*, \label{aa1} \\
a_2&=&\frac{\Delta +\ell }{(\Delta +\ell +1) (\Delta -R+\ell ) (\Delta +R+\ell )}
\left(\la_{\Phi_1\Phi_2\cO}^{(1)}+  \frac{r (\Delta -R+\ell )}{2 (\Delta +\ell )}\la_{\Phi_1\Phi_2\cO}^{(0)} \right)  \nonumber \\
&&\times\left((\la_{\Phi_4^\dagger\Phi_3^\dagger\cO}^{(1)})^* -\frac{\tl r (\Delta -R+\ell )}{2 (\Delta +\ell)}
(\la_{\Phi_4^\dagger\Phi_3^\dagger\cO}^{(0)})^*\right), \label{aa2} \\
a_3&=&\frac{\ell +2-\Delta }{(-\Delta +\ell +1) (-\Delta -R+\ell +2) (-\Delta +R+\ell +2)}  \nonumber \\
&&\times\left(\la_{\Phi_1\Phi_2\cO}^{(1)}+ \frac{\ell+1}{\ell}\la_{\Phi_1\Phi_2\cO}^{(3)}+\frac{r (-\Delta +R+\ell +2)}{2 (-\Delta +\ell +2)}\la_{\Phi_1\Phi_2\cO}^{(0)} \right) \nonumber\\
&&\times\left((\la_{\Phi_4^\dagger\Phi_3^\dagger\cO}^{(1)})^* +\frac{\ell+1}{\ell}(\la_{\Phi_4^\dagger\Phi_3^\dagger\cO}^{(3)})^*-\frac{\tl r (-\Delta +R+\ell +2)}{2 (-\Delta +\ell +2)}(\la_{\Phi_4^\dagger\Phi_3^\dagger\cO}^{(0)})^*\right),  \label{aa3}\\
a_4 &=&\frac{4 (\Delta -1)^2 (-\Delta +\ell +2) (\Delta +\ell )}{\Delta ^2 (\ell +1-\Delta) (\Delta +\ell +1) (\ell +2-R-\Delta) (\ell +2+R-\Delta) (\Delta -R+\ell ) (\Delta +R+\ell )} \times\nonumber\\
&&\left[ -\frac{(\Delta -R+\ell ) \left(R \left(\ell  (\ell +2)-\Delta  \left(\Delta +r^2-2\right)\right)+(\ell +2-\Delta) \left((\Delta +\ell )^2-\Delta  r^2\right)\right)}{8 (\Delta -1) (\ell +2-\Delta) (\Delta +\ell )}
\la_{\Phi_1\Phi_2\cO}^{(0)} \right. \nonumber \\
&&\left.
+ \la_{\Phi_1\Phi_2\cO}^{(2)}+\frac{r (\Delta  (R+2-\Delta)+\ell  (\ell +2))}{2 (\ell +2-\Delta ) (\Delta +\ell )}\la_{\Phi_1\Phi_2\cO}^{(1)}+\frac{r (\Delta -R+\ell )}{4 (\Delta +\ell )}\la_{\Phi_1\Phi_2\cO}^{(1)}\right]\times  \nonumber \\
&&\left[ -\frac{(\Delta -R+\ell ) \left(R \left(\ell  (\ell +2)-\Delta  \left(\Delta +r^2-2\right)\right)+(\ell +2-\Delta) \left((\Delta +\ell )^2-\Delta  r^2\right)\right)}{8 (\Delta -1) (\ell +2-\Delta ) (\Delta +\ell )}
(\la_{\Phi_4^\dagger\Phi_3^\dagger\cO}^{(0)})^*  \right. \nonumber \\
&&\left.
+ (\la_{\Phi_4^\dagger\Phi_3^\dagger\cO}^{(2)})^*-\frac{\tl r (\Delta  (R+2-\Delta)+\ell  (\ell +2))}{2 (\ell +2-\Delta) (\Delta +\ell )}(\la_{\Phi_4^\dagger\Phi_3^\dagger\cO}^{(1)})^*-\frac{\tl r (\Delta -R+\ell )}{4 (\Delta +\ell )}(\la_{\Phi_4^\dagger\Phi_3^\dagger\cO}^{(3)})^*\right].
\label{aa4}
\eea
}

\end{document}